\pdfoutput=1

\PassOptionsToPackage{table,xcdraw}{xcolor}
\documentclass[sigconf, nonacm]{acmart}
\usepackage{xcolor}
\usepackage{verbatim}
\usepackage{listings}
\newcommand{\revision}[1]{#1}
\newcommand{\old}[1]{}
\newcommand{\change}[2]{#2}

\newcommand{\perf}[1]{{\textbf{#1}}\ensuremath{\times}}
\newcommand{\SpeedupSTDuckTPCH}{\perf{11.17}}
\newcommand{\SpeedupSTDuckCEB}{\perf{45.33}}
\newcommand{\SpeedupSTUmbraTPCH}{\perf{7.24}}
\newcommand{\SpeedupSTUmbraCEB}{\perf{9.56}}

\newcommand{\SpeedupMTDuckTPCH}{\perf{7.65}}
\newcommand{\SpeedupMTDuckCEB}{\perf{23.97}}
\newcommand{\SpeedupMTUmbraTPCH}{\perf{6.12}}
\newcommand{\SpeedupMTUmbraCEB}{\perf{1.87}}

\newcommand{\revisionimage}[1]{#1}

\newcommand\vldbdoi{XX.XX/XXX.XX}
\newcommand\vldbpages{XXX-XXX}
\newcommand\vldbvolume{19}
\newcommand\vldbissue{11}
\newcommand\vldbyear{2026}
\newcommand\vldbauthors{\authors}
\newcommand\vldbtitle{\shorttitle} 
\newcommand\vldbavailabilityurl{https://github.com/DataManagementLab/BespokeOLAP}
\newcommand\vldbpagestyle{empty}

\usepackage{hyperref}
\usepackage{cleveref}
\usepackage{caption}
\usepackage{graphicx}
\usepackage{subcaption}
\usepackage[disable]{todonotes}
\usepackage{pifont}
\usepackage[table,xcdraw]{xcolor}
\usepackage{pbalance}
\usepackage{multicol}
\usepackage[normalem]{ulem}
\usepackage{multirow}
\renewcommand{\paragraph}[1]{\noindent\textbf{#1}$\:$}

\begin{document}

\title[Bespoke OLAP: Synthesizing Workload-Specific One-size-fits-one Database Engines]{Bespoke OLAP: Synthesizing Workload-Specific\\One-size-fits-one Database Engines}

%Bespoke OLAP: Synthesizing Workload-Specific Database Engines on Demand
%Beyond General-Purpose: Synthesizing OLAP Engines That Fit Exactly One Workload
%%
%% The "author" command and its associated commands are used to define the authors and their affiliations.
\author{Johannes Wehrstein}
\orcid{0000-0002-7152-8959}
\affiliation{%
  \institution{Technical University of Darmstadt}
  % \city{Darmstadt}
  % \country{Germany}
}
% \email{johannes.wehrstein@tu-darmstadt.de}
\author{Timo Eckmann}
\affiliation{%
  \institution{Technical University of Darmstadt}
  % \city{Darmstadt}
  % \country{Germany}
}
% \email{timo.eckmann@tu-darmstadt.de}
\author{Matthias Jasny}
\authornote{Work done while at Technical University of Darmstadt.}
\affiliation{%
  \institution{Microsoft Gray-Systems-Lab}
  % \city{Darmstadt}
  % \country{Germany}
}
% \email{matthias.jasny@tu-darmstadt.de}
\author{Carsten Binnig}
\orcid{0000-0002-2744-7836}
\affiliation{%
  \institution{Technical University of Darmstadt \& DFKI \& hessian.AI}
  % \city{Darmstadt}
  % \country{Germany}
}
% \email{carsten.binnig@tu-darmstadt.de}

%%
%% The abstract is a short summary of the work to be presented in the
%% article.
\begin{abstract}
\change{}{Modern OLAP engines support arbitrary analytical workloads, but this flexibility incurs overhead from runtime schema interpretation, generic data representations, and abstraction layers, even in compiled-query systems. Workload-specific engines can eliminate these costs and exploit specialized data structures and algorithms for higher performance, yet have historically been too expensive to build manually. Recent advances in LLM-based code synthesis challenge this tradeoff, but naive prompting does not produce correct or efficient engines due to deep architectural dependencies and the need for systematic refinement. We present \emph{Bespoke OLAP}, a fully autonomous synthesis pipeline that constructs high-performance OLAP engines tailored to a target workload through iterative performance evaluation and automated validation. Bespoke OLAP generates engines from scratch within minutes to hours and achieves order-of-magnitude speedups over DuckDB and Umbra, demonstrating that the generality tax extends beyond query compilation to storage layout and algorithmic design.}
\end{abstract}

\maketitle
% Clean PDF metadata title (set after \maketitle, which acmart uses to write pdftitle;
% the \\ in \title otherwise drops the space, yielding "Workload-SpecificOne-size").
\hypersetup{pdftitle={Bespoke OLAP: Synthesizing Workload-Specific One-size-fits-one Database Engines}}

%%% do not modify the following VLDB block %%
%%% VLDB block start %%%
\pagestyle{\vldbpagestyle}
\begingroup\small\noindent\raggedright\textbf{PVLDB Reference Format:}\\
\vldbauthors. \vldbtitle. PVLDB, \vldbvolume(\vldbissue): \vldbpages, \vldbyear.\\
\href{https://doi.org/\vldbdoi}{doi:\vldbdoi}
\endgroup
\begingroup
\renewcommand\thefootnote{}\footnote{\noindent E-mail:~\texttt{firstname.lastname@tu-darmstadt.de}\\
This work is licensed under the Creative Commons BY-NC-ND 4.0 International License. Visit \url{https://creativecommons.org/licenses/by-nc-nd/4.0/} to view a copy of this license. For any use beyond those covered by this license, obtain permission by emailing \href{mailto:info@vldb.org}{info@vldb.org}. Copyright is held by the owner/author(s). Publication rights licensed to the VLDB Endowment. \\
\raggedright Proceedings of the VLDB Endowment, Vol. \vldbvolume, No. \vldbissue\ %
ISSN 2150-8097. \\
\href{https://doi.org/\vldbdoi}{doi:\vldbdoi} \\
}\addtocounter{footnote}{-1}\endgroup

\ifdefempty{\vldbavailabilityurl}{}{
\vspace{.3cm}
\begingroup\small\noindent\raggedright\textbf{PVLDB Artifact Availability:}\\
The source code, data, and/or other artifacts have been made available at \url{\vldbavailabilityurl}.
\endgroup
}
%%% VLDB block end %%%

\section{Introduction}
\label{sec:intro}

\begin{figure}
\captionsetup{aboveskip=0.0ex,belowskip=0.0ex}
\captionsetup[subfigure]{aboveskip=0.0ex,belowskip=0.0ex}
\centering
\revisionimage{
\includegraphics[width=\linewidth]{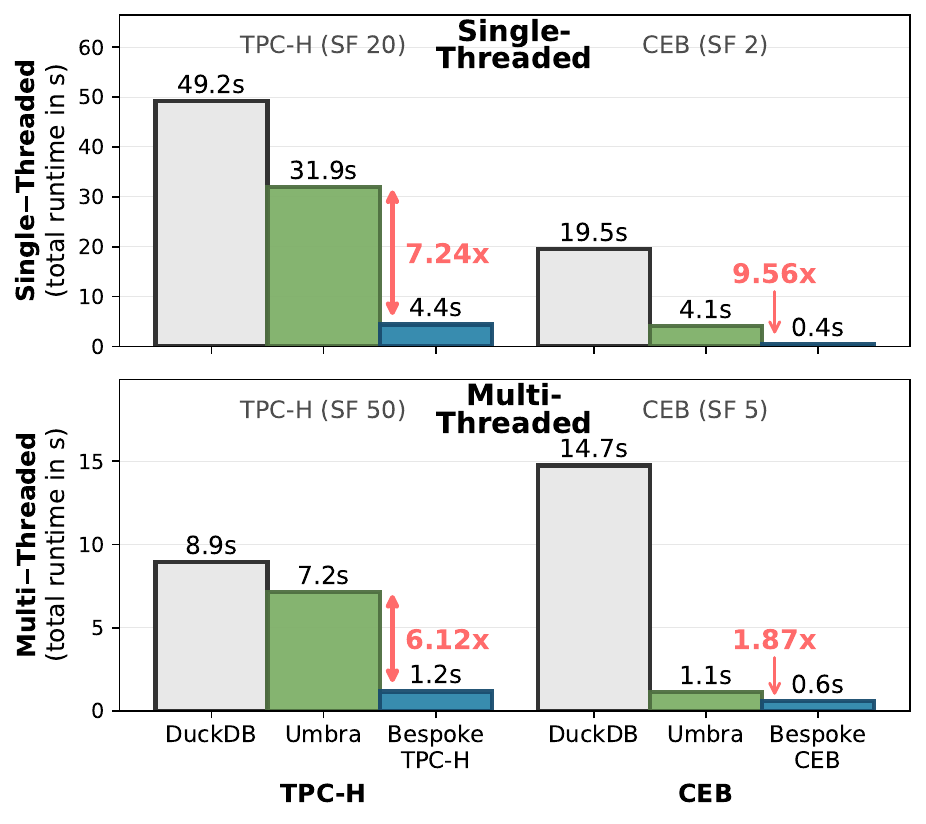}
}
\Description{}
%\vspace{-3ex}
%\caption{Synthesized Engines Outperform a Decade of Engineering.
%The synthesized Bespoke TPC-H and CEB engines achieve an
%11.78$\times$ / 9.76$\times$ lower total runtime and a 16.40$\times$ / 4.66$\times$ median per-query
%speedup over DuckDB (v1.4.4) at scale factor~20 (TPC-H)/2 (CEB).}
%\caption{Synthesized Bespoke TPC-H and CEB engines exceed the performance of a decade of engineering, achieving 11.78$\times$/9.76$\times$ lower total runtime and 16.40$\times$/4.66$\times$ median per-query speedups over DuckDB.}
\caption{\change{Synthesized Bespoke engines exceed the performance of decades of engineering, achieving 7.24$\times$ and 9.56$\times$ lower total runtime on TPC-H and CEB, respectively, over Umbra.}{Synthesized Bespoke engines exceed the performance of a decade of engineering, achieving \SpeedupSTDuckTPCH{} and \SpeedupSTDuckCEB{} lower total runtime on TPC-H and CEB over DuckDB in single-threaded execution, outperforming even Umbra, which uses compiled query execution (\SpeedupSTUmbraTPCH{} on TPC-H and \SpeedupSTUmbraCEB{} on CEB). Similar speedups are visible in multi-threaded execution.}}
%\vspace{-3ex}
\label{fig:speedup}
\end{figure}

\paragraph{OLAP Engines Carry an Inherent Performance Tax.}
It is well established that one size does not fit all~\cite{stonebraker2005onesize}. 
Over the past decades, this insight has led to workload-class-specific systems, including columnar OLAP engines such as DuckDB~\cite{raasveldt2019duckdb} and HyPer~\cite{kemper2011hyper}, which depart significantly from traditional row-store architectures. 
Yet even these systems follow a refined version of the ``one-size-fits-all'' philosophy: they are designed for \emph{arbitrary} OLAP workloads within the relational model. 
They must support any schema, any valid SQL query, and a broad spectrum of access patterns.
This flexibility introduces an inherent performance tax. 
Schemas are interpreted at runtime, tuple layouts remain generic, and data structures are chosen to accommodate unknown future queries. 
While query compilation reduces interpretation overhead for individual queries~\cite{neumann2011efficiently}, engines still operate within a generic storage and execution layer where all kinds of SQL queries can be executed, and thus general-purpose algorithms and data structures for storing tables are used. 
Importantly, this overhead is not accidental or due to poor engineering;
it is the unavoidable price of generality.

\paragraph{Generality is an Overhead.}
Crucially, many real-world OLAP deployments do not require such flexibility. 
Enterprise data warehouses frequently execute stable suites of parameterized query templates. 
Recent workload analyses confirm extremely high query repetition rates in practice~\cite{van2024tpc,marcus2023learned,wehrstein2025redbench}.
In these settings, the ability to run arbitrary ad hoc queries is rarely exercised, yet its cost is incurred on every execution as a general-purpose OLAP engine is used.
Instead, using a database engine designed for a specific workload with a fixed schema and a known set of queries eliminates the overhead of generality by construction. 
More precisely, schema interpretation disappears entirely, storage layouts can be organized around actual access patterns, and encoding decisions can be tailored to observed value distributions. 
Furthermore, all queries can be compiled into code that directly operates on workload-specific data structures rather than on generic relational operators. 
As a consequence, entire subsystems such as dynamic planning, generic operator selection, and fallback code paths can be removed from the runtime path.
The result is not a better general-purpose engine, but a different artifact: an engine correct and maximally optimized for a single workload.
In short, the optimal OLAP engine for a workload is what we call a bespoke ``one-size-fits-one'' engine.

\paragraph{Why Bespoke Engines Have Been Economically Out of Reach.}
If the performance benefits are so clear, why are bespoke OLAP engines not common practice?
The obstacle has never been conceptual feasibility but rather economics.
Building a high-performance DBMS requires years of engineering by expert teams. 
Storage layout, indexing, execution strategies, memory management, and testing infrastructure are deeply interdependent; even small design decisions propagate throughout the system. 
Constructing a bespoke engine is therefore comparable in cost to building a new database system.
Rare examples such as TigerBeetle~\cite{tigerbeetle2024} demonstrate that extreme specialization can pay off when the target workload is narrow yet widely shared. 
However, most analytical workloads are organization-specific. 
In a world with potentially millions of distinct workloads, manually building an engine for each workload requires millions of engines. 
Engineering effort scales linearly with workload diversity, rendering a manual bespoke design economically infeasible. 
Consequently, as of today, general-purpose engines are used despite the clear performance benefits of bespoke engines.
The performance tax of generality is therefore paid indefinitely, not because specialization does not help, but because manual specialization does not scale.

\paragraph{LLM-Guided Synthesis Makes Bespoke Engines Practical.}
Recent advances in AI-based code generation fundamentally change this tradeoff. 
Large language model (LLM) coding agents have demonstrated the ability to generate substantial, non-trivial software artifacts at low cost and within hours rather than years. 
This raises a natural question: can we automatically synthesize a bespoke OLAP engine for a given workload specification?
In this paper, we investigate this question.
Given a workload specification consisting of query templates, and performance objectives, our LLM-driven pipeline generates a complete, workload-specific analytical engine in minutes to a few hours, at a cost of only a few dollars. 
Importantly, the generated engine is not a configured instance of an existing system, instead, it is constructed from scratch for one workload and aggressively optimized for it. 
The resulting performance gains are substantial. 
Figure~\ref{fig:speedup} shows that synthesized engines for TPC-H and CEB achieve double-digit reductions in total runtime over DuckDB and Umbra.
These improvements reflect structural overhead imposed by generality rather than incremental implementation differences and demonstrate that a significant fraction of modern OLAP cost arises from remaining general-purpose design.

%\paragraph{The Scientific Challenge: Complex System Synthesis.}
\paragraph{The Challenge of Synthesizing OLAP Engines.}
Naively prompting does not work. 
%In theory, database engines should be well-suited to be automatically synthesized because a database always consists of the same fundamental components such as storage, data access, or operator processing. 
Initially, we tried to simply use a coding agent to synthesize a bespoke engine. However, we ran into several problems. Firstly, correctness was not consistently reached and it took intensive human guiding and many iterations to finally generate a seemingly bespoke engine. Additionally, even if the coding agent produced a correct execution strategy it was not performant. This is mainly due to database engines being complex systems of deeply interdependent components as storage formats constrain operator design, execution strategies depend on data layout, and optimizations in one layer can introduce regressions in another. 
%However, naively prompting an LLM to ``generate a full database engine'' for a given workload does not work even with the most advanced coding models - and will likely not work as important signals such as performance feedback are missing. 
%Furthermore, database engines are complex systems of deeply interdependent components: storage formats constrain operator design, execution strategies depend on data layout, and optimizations in one layer can introduce regressions in another. 
Therefore, in our early testing, full OLAP engines that are naively generated by LLMs often fail to compile, violate correctness, or exhibit fragile performance across queries.
The core scientific problem is therefore not code generation per se, but structured system synthesis of complex systems, which maintains correctness while aggressively optimizing performance. 
As such, without principled structure and validation, LLM-driven development quickly degenerates into uncontrolled modification rather than systematic optimization.

\paragraph{Our Contribution: A Pipeline for Bespoke Engine Synthesis.}
%Database engines are well-suited to be automatically synthesized because a database always consists of the same fundamental components such as storage, data access, operator processing or 
This paper presents a synthesis pipeline that meets this challenge. 
First, generation proceeds incrementally: the agent constructs and validates a functional storage layer before introducing execution logic, and queries are synthesized one by one to localize complexity and prevent error propagation. 
Second, correctness and performance optimization are strictly separated. 
We first establish a validated, correct baseline and then iteratively optimize for performance without sacrificing correctness.
However, incrementally changing code and then running it to test correctness and performance comes at a high cost.
To make this optimization loop practical, we thus also contribute infrastructure for system synthesis that enables live hotpatching of database engines. 
This allows the synthesis process to modify a running engine and immediately observe performance effects, reducing validation and benchmarking time from minutes to seconds. 
Candidate implementations are benchmarked on the target data (downscaled) and queries, while fine-grained runtime profiles identify bottlenecks that guide subsequent synthesis steps. 
Unlike classical database engines, which rely on cost models to predict optimal strategies at runtime, our approach can thus directly measure alternatives during synthesis and hard-code the fastest one into the final artifact.

\paragraph{Outline.}
In Section~\ref{sec:overview}, we provide a more detailed overview of Bespoke OLAP synthesis. 
Section~\ref{sec:approach} then presents the complete design of our synthesis pipeline, followed by Section~\ref{sec:system}, which describes the infrastructure for system synthesis, which involves live hotpatching of database engines, regression tracking, and empirically driven optimization. 
Section~\ref{sec:evaluation} provides a comprehensive evaluation on TPC-H~\cite{tpch} and CEB~\cite{flowlossceb}, demonstrating consistent and substantial speedups over DuckDB across all query templates. 
We then discuss related work in Section \ref{sec:related_work}, where we distinguish our approach from query compilation, auto-tuning, and learned components.
Finally, to conclude, we summarize the contributions and outline research directions.

\begin{figure*}[t]
\captionsetup{aboveskip=0.0ex,belowskip=0.0ex}
\captionsetup[subfigure]{aboveskip=0.0ex,belowskip=0.0ex}
    \centering
    \includegraphics[width=\linewidth]{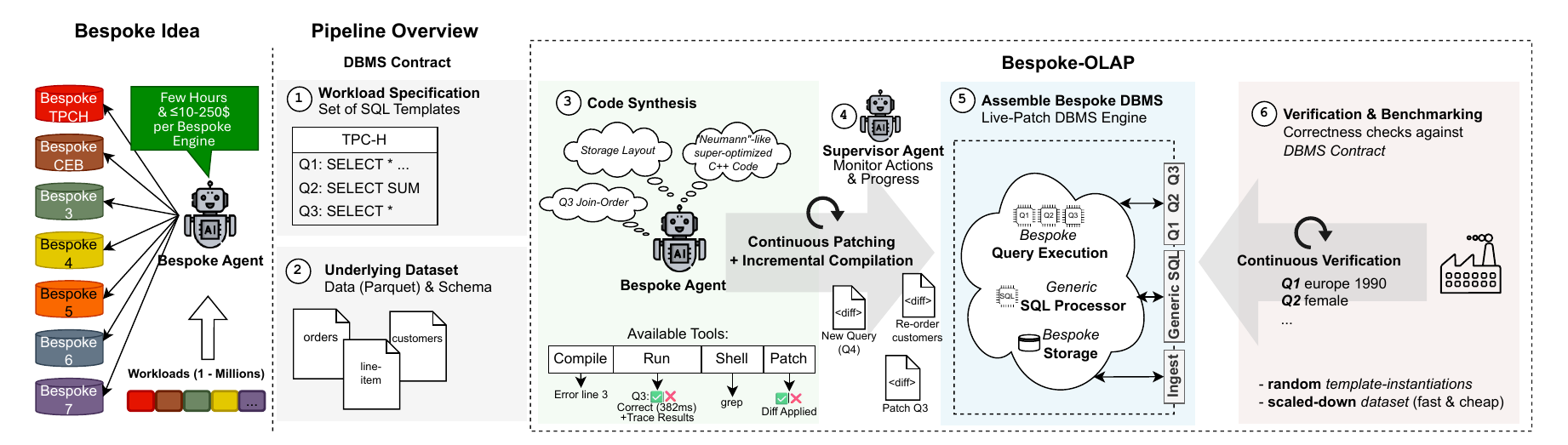}
    \Description{}
    \caption{Overview of the Bespoke OLAP idea: one engine is generated per workload (left), and the pipeline to synthesize one engine (right). The pipeline works as follows:
    Based on the DBMS Contract (SQL Query Templates \textcircled{1} and Dataset \textcircled{2}), the Bespoke Agent iteratively generates the query execution engine\textcircled{3}, checked by the supervisor agent \textcircled{4}, compiles, deploys\textcircled{5}, and validates it\textcircled{6}, and uses performance and correctness results to guide further code generation.
    At runtime, the generated bespoke engine exposes interfaces for ingest, for the query workload, and even ad-hoc queries, which are routed to a generic SQL processor \textcircled{5}.}
    \label{fig:approach_overview}
    \vspace{-2ex}
\end{figure*}

\section{Overview of Bespoke OLAP}
\label{sec:overview}

In this section, we first present an overview of Bespoke OLAP.
Afterwards, we also discuss the scope of this work. 

\subsection{Synthesizing Bespoke Engines}
Figure~\ref{fig:approach_overview} provides an overview of our approach.
At the core is the so-called Bespoke Agent, which implements our synthesis process and generates the code for the Bespoke OLAP engines.

\paragraph{A Bespoke Engine is Defined by the Workload.} \label{subsec:synthBespokeEngines}
The starting point of bespoke synthesis is a formal commitment of a user who wants to synthesize an engine, which we call the \emph{DBMS contract}. 
This contract consists of two parts: 
The first part of the contract, as shown in Figure \ref{fig:approach_overview} at step \textcircled{1}, is a workload specification comprising the database schema together with a set of query templates and their parameterization ranges, which defines exactly what the engine must be capable of executing. 
The second part of the contract is the underlying dataset (\textcircled{2} in Figure~\ref{fig:approach_overview}), provided as a collection of Parquet files, from which the engine derives its physical storage layout during an initial ingestion phase. 
Together, these two components define the complete design space available for a Bespoke engine, and every decision in step \textcircled{3} in Figure~\ref{fig:approach_overview} is grounded in what the contract reveals about the data and the queries. \revision{Naturally, a bespoke engine can only be generated for workloads whose query templates and correct query results are known in advance.}
Importantly, generation is done incrementally as described below, and the generated bespoke DBMS is verified and benchmarked in every iteration (see \textcircled{6}) to guarantee correctness and performance and guide the code generation.

\paragraph{Bounded Scope is the Strength of Bespoke OLAP.}
The contract also serves as a boundary; i.e., queries that fall outside the specification are explicitly not supported by the bespoke engine and are instead handled by a fallback database engine (i.e., a general-purpose DBMS or a generated general SQL processor over the Bespoke Storage) as shown in \textcircled{5} in Figure~\ref{fig:approach_overview}. 
As a prime application of Bespoke OLAP, we see typical warehousing dashboards that execute a fixed set of queries.
Additionally, any OLAP workload with repeating queries, which are highly common as discussed before, are an application for Bespoke OLAP.
Importantly, the bounded scope of a Bespoke OLAP engine is not a limitation but the very source of the engine's performance advantage, since it is precisely because the workload is fixed that the pipeline can eliminate every abstraction that exists solely to handle cases the workload will never present. 
%On the interface side, the pipeline enforces a minimal and fixed contract with the outside world. 
Based on the two given inputs, namely the database and query templates, the full system is generated by the Bespoke Agent, including the storage layout and data ingestion code, the data structures used to represent each table, the execution code for each query, etc..
Therefore, the engine that emerges is not a configured instance of any existing system but a purpose-built artifact whose entire internals are shaped by the contract with its workload.

%\paragraph{Regeneration and General-Purpose Fallback Engine}
%\todo{partially covered in \Cref{subsec:changing_wl}}

%\subsection{Insights and Discussion}
%\paragraph{A Single Prompt Is Not Sufficient.}
%One may believe that it is already possible to generate a bespoke DBMS with a naive "generate a database engine" prompt. However, our early experiments have shown that this is not feasible. The problems we encountered were 

\paragraph{Synthesis Requires Both: Pipeline and Infrastructure.}
With the contract defined and the scope fixed, the remaining question is how to reliably guide an LLM agent in constructing an optimized Bespoke engine. 
As we describe in detail in Section~\ref{sec:approach}, this requires two things working together. 
(1) The first component is the synthesis pipeline (also called the agent loop in this paper) that breaks synthesis into well-defined stages, moving from storage layout planning through basic query implementation and into a multi-round optimization phase driven entirely by empirical measurement rather than cost estimation. 
(2) The second component is a dedicated infrastructure that keeps the development loop fast, allowing the Bespoke Agent to test the synthesized engine quickly for performance and correctness.
Here, we developed a system for system generation that includes techniques for hotpatching database engines, protection against regressions through external rollback, and ensures correctness through continuous fuzzy testing.\\
Together, these two components make reliable synthesis possible, and Section~\ref{sec:approach} and Section \ref{sec:system} describe both in detail.

\paragraph{How to Support Workload Changes and Ad-hoc Queries?}
Because a bespoke engine is specialized for a fixed workload, an important question is what happens when previously unseen queries arrive or when the workload shifts over time. 
We address this at two levels. 
For occasional ad-hoc queries that fall outside the contract, the engine can route execution through a general-purpose SQL interface that operates on the same underlying storage layer, thereby providing correct results without workload-specific optimization. 
For more substantial workload changes, the engine can be resynthesized entirely, and because synthesis completes in minutes to hours at the cost of a few dollars, regeneration is a practical response to workload drift rather than a theoretical fallback and can even be done on a high frequency, such as every night during batch loading new data. 
Resynthesis can also either reuse the existing storage layer to avoid re-ingesting the data if the schema does not change or include a full reorganization of the physical layout to optimally support the new workload. 
We describe both mechanisms in more detail in Section~\ref{subsec:changing_wl}.

\subsection{Scope of this Work}

\paragraph{C++ as Language for the Engine}
Our pipeline is language agnostic. However, an early design decision we made in the pipeline for our experiments is the target programming language, as it determines the quality of compiler feedback, the available optimization headroom, and how reliably agents can reason about generated code.
We chose C++ because it is well understood by current LLMs, enabling idiomatic and performant code generation. 
It provides fine-grained control over memory layout and low-level execution, where workload-specific optimizations arise; higher-level languages would obscure these physical decisions. 
% Additionally, its compilation errors are precise and localizable, offering clear feedback when code is incorrect.
Additionally, its tooling and compilation infrastructure offers clear feedback to the LLM.

\paragraph{In-Memory OLAP Engines.} To show that bespoke system generation is possible and see the order in which we can expect benefits, the synthesis pipeline described in this paper limits the design to in-memory engines. This eliminates the need to generate code for buffer management, page eviction, and I/O scheduling from the design space, allowing the agent to focus its optimization effort on data structure design and computation rather than on data movement between storage tiers. \change{Similarly, we decided to restrict execution to a single thread, removing parallelization from the space of decisions the Bespoke Agent must make, ensuring that performance differences between the synthesized engine and the reference system reflect specialization of algorithms rather than parallelism. However, it can be straightforwardly parallelized after single-threaded execution achieves high performance using techniques such as morsel-driven execution~\cite{leis2014morsel}.}{We evaluate both single-threaded and multi-threaded execution. Single-threaded results isolate the contribution of specialization from parallelism, and multi-threaded results demonstrate that the synthesized engines scale effectively with additional cores.} These are deliberate restrictions rather than fundamental limitations of the bespoke synthesis approach to see how well bespoke adaptation on the full system level can help. Within this scope, we demonstrate that bespoke synthesis can produce engines that substantially outperform highly optimized general-purpose systems, which is the central claim this paper sets out to establish. \change{Extending the pipeline to disk-resident data and multi-threaded execution introduces new opportunities (e.g., bespoke storage formats) that we will address in future work. However, we believe that neither of these extensions changes our core synthesis methodology, which we describe in detail next.}{Extending the pipeline to disk-resident data introduces new challenges that we mention as future work in Section~\ref{sec:conclusion}. All generated engines build on the same core synthesis methodology, which we describe in detail next.}

\begin{figure}
\captionsetup{aboveskip=0.0ex,belowskip=0.0ex}
\captionsetup[subfigure]{aboveskip=0.0ex,belowskip=0.0ex}
    \centering
    \includegraphics[width=\linewidth]{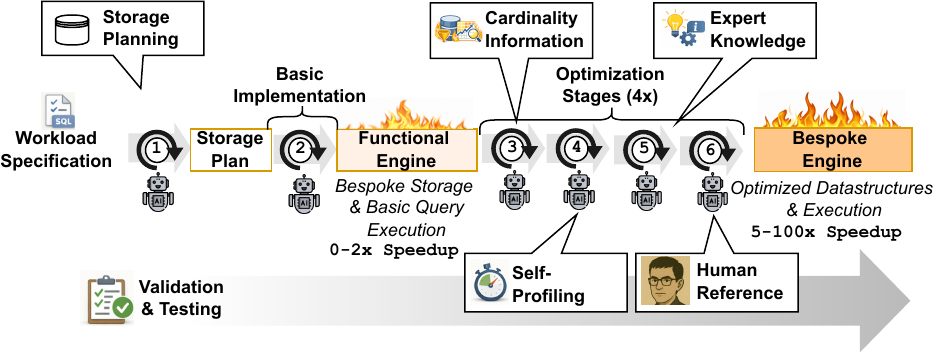}
    \Description{}
    \caption{Steps to produce a Bespoke Engine. Each step is executed in loops, with validation to check whether the step's goal is achieved.
    \textcircled{1} produces a storage plan for the specified workload.
    \textcircled{2} produces a functional engine based on the storage plan coupled with continuous validation and testing to steer the iterative coding process (see arrow below).
    The optimization loop produces an optimized engine iteratively by leveraging additional information such as actual summaries of the data \textcircled{3}, self-profiling \textcircled{4}, expert knowledge from the DBMS research literature\textcircled{5}, and taking the perspective of expert database engineers \textcircled{6}.}
    \label{fig:optim_stages}
\end{figure}

\section{The Generation Pipeline}
\label{sec:approach}

In this section, we describe the generation pipeline for synthesizing bespoke OLAP engines, walking through each stage from storage planning through basic query implementation to the four steps of optimization that produce the final bespoke engine. 
An overview of all the synthesis stages is illustrated in Figure~\ref{fig:optim_stages}.

\subsection{Overview of Pipeline}

\paragraph{The Environment of the Bespoke Agent.}
Throughout the entire synthesis process, the Bespoke Agent interacts with the environment exclusively through four tools shown in \Cref{fig:approach_overview} \textcircled{3}. 
The first is a \emph{compiler tool} that compiles the current state of the codebase and returns any errors with precise file and line information, giving the agent actionable feedback whenever a change introduces a syntactic or type error. 
The second is a \emph{benchmark tool} that executes the current engine against the workload, reports correctness by comparing results against DuckDB across a set of parameter instantiations, and returns runtime measurements for each query. 
The third is a \emph{shell tool} that allows the agent to inspect the underlying dataset, examine intermediate values during execution, and explore the structure of the codebase. 
The fourth is a \emph{patch tool} through which the agent submits all code changes as structured patches rather than full file rewrites. 
This last tool is the interface through which hotpatching operates, which we explain more in Section \ref{sec:system}.
%As we discuss in Section \ref{sec:system}, our system for system generation intercepts the patch, recompiles only the affected component, and swaps it into the running engine. 
All four tools together define the complete boundary between the agent and the system it is building, and the agent has no mechanism to interact with the engine other than through them.

\paragraph{Synthesis Begins With Storage Layout Planning.}
The first stage of the agent loop is storage layout planning, shown in \Cref{fig:optim_stages} \textcircled{1}, and it is deliberately the only stage in which no execution code is written.
Before the agent generates a single query function, it uses the shell tool to examine the underlying dataset in detail and cross-references its findings with the query templates in the workload specification. 
It identifies which columns appear in predicates, which attributes serve as join keys between tables, which columns are accessed together by multiple queries, and how data is distributed across the relevant attributes. 
Based on this analysis, the agent decides on the physical storage layout for each table (called storage plan), selecting sort orders, encoding schemes, and auxiliary data structures that are optimal for the specific access patterns the workload presents.
This planning phase is completed and committed before any code is written. The reason for this strict ordering is that storage decisions affect every query in the workload simultaneously. 
Allowing the agent to revisit storage decisions during query implementation would introduce cross-query dependencies that are very difficult to manage reliably and would undermine the correctness of already-validated queries.
Hence, changes to the storage layout should be kept to a minimum to avoid this overhead.
Once the initial storage layout is determined, the agent proceeds to the second stage and generates the corresponding C++ structs \textcircled{2} for storing the data together with the ingestion code that loads data, transforms them according to the storage layout, and populates the in-memory storage representation. 
After ingestion, the code is compiled, and the database process is started, which then remains running until the complete synthesis finishes.

\paragraph{The Base Ingredients: Correctness \& Tracing.} \label{para:Correctness&Tracing}
With a running engine and a stable storage layer in place, the Bespoke Agent next proceeds to implement a dedicated execution function for each SQL query template in the workload specification. 
At this stage, the goal is to generate an engine that correctly executes all queries, rather than optimizing for performance. 
%The agent implements each query using straightforward algorithms, validates the result with the validate tool against DuckDB, and moves on to the next query. 
Correctness is checked by comparing query results against an existing engine such as DuckDB.
This basic implementation phase establishes the known-good baseline that all subsequent optimization will be measured against. 
Once the full set of query templates executes correctly, the agent performs two preparatory steps before optimization begins. 
The first is the addition of CPU core pinning, for which we provide C++ helper functions at specific positions in the execution code, to ensure reliable benchmarking results.
The second preparatory step is the installation of tracing functionality, which includes lightweight performance counters, loop iteration trackers, and timestamps that the agent autonomously injects into the execution code wherever it finds additional statistics meaningful.
These agent-injected tracing steps will later be used by the agent to verify its implementation decisions and optimize the query code for performance.
All tracing code is guarded by a preprocessor directive and can be disabled by the agent, such that it imposes no overhead for the final engine. 
\paragraph{Optimization Proceeds in a \revision{Supervised} Loop.}\label{subsec:supervisedLoop}
With the instrumentation in place, the Bespoke Agent enters the optimization phase, which proceeds through four rounds of increasing sophistication (shown in the middle of Figure \ref{fig:optim_stages} \textcircled{3}-\textcircled{6}).
\revision{A dedicated supervisor agent observes each turn of the optimization loop, verifying that the agent invoked the appropriate tools, completed the assigned task, and introduced no regressions. When the supervisor detects gaps or errors, it injects targeted corrective feedback directly into the agent's context; otherwise it approves the turn and the loop advances. The supervisor is stage-aware: it knows the sequence of upcoming stages and refrains from pushing the agent to address concerns that are reserved for a later round.}
In the first round \textcircled{3}, the agent receives signals from executing one randomly sampled query plan per query template from DuckDB, annotated with automatically retrieved actual cardinalities.
This additional information about cardinalities for one possible instantiation of a query template is the starting point for the agent to iteratively try and empirically evaluate optimization decisions, such as join order and operator implementations, as we will discuss later.
Bootstrapping this empirical optimization loop with the cardinalities of one sample instantiation per query-template allows the agent to skip highly suboptimal join-orders from the search space right from the start, speeding up the optimization process.

In the second round \textcircled{4}, the agent activates the tracing framework \textcircled{4} and analyzes the instrumentation output it produces. 
It examines where time is spent within each query function, identifies loop bodies that dominate execution time, and pinpoints operators whose implementation does not match the scale of intermediate results.
This self-directed profiling allows the agent to localize bottlenecks precisely rather than making speculative changes to the entire function. 
In the third round \textcircled{5}, the agent draws on distilled expert knowledge \textcircled{5} from the database research literature that we have curated and provided as part of the synthesis environment. 
This knowledge covers established optimization techniques from decades of systems research, focused on keeping data access sequential, reducing cache misses and copies, hoisting loop invariants, minimizing branches, and removing expensive work from inner loops. 
It also recommends choosing data-appropriate algorithms and containers, pre-sizing outputs, using direct parsing and constant-time lookups, and helping the compiler and hardware with aliasing hints, inlining, and SIMD-friendly code\footnote{The complete expert-knowledge file can be found \href{ https://github.com/DataManagementLab/BespokeOLAP/blob/main/conversations/prompts/expert_knowledge.txt}{\textit{here}}.}.
The agent applies these techniques selectively to the bottlenecks it identified through tracing and validates each application against the benchmark tool to confirm that the expected improvement materialized. 
In the fourth and final round \textcircled{6}, the agent adopts the perspective of an expert database engineer and performs a holistic review of each query function, looking for any remaining inefficiency that the previous rounds did not address. 
It treats the tracing output as its primary guide and reasons about the generated code the way a systems engineer would, questioning every allocation, every branch, and every data movement that cannot be justified by the workload's access patterns. 
Changes made in this round are validated and checkpointed exactly as in earlier rounds, so that any regression is detected immediately and reverted before it can affect other queries.

\paragraph{No cost modeling, but empirically finding join-orders.}
Since query templates are parameterized, different instantiations can yield vastly different intermediate cardinalities and thus different optimal join orders, something a static cost model in a generic DBMS would have to predict. In practice, however, accurate cost models for arbitrary operator implementations do not exist. Our Bespoke Agent therefore sidesteps cost modeling entirely and discovers effective join orders empirically: it proposes an ordering, benchmarks it on sampled instantiations, and iteratively refines the design based on observed runtimes. The cardinality information with an optimal plan, as provided in Figure \ref{fig:optim_stages} \textcircled{3}, eliminates suboptimal join orders early, narrowing the search space before deeper exploration. When parameter instantiations consistently favor different join orders, the agent either (1) selects a single robust plan that performs well across samples or (2) generates separate specialized implementations for distinct regions of the parameter space, tailoring the code to specific argument profiles. This iterative propose-and-evaluate loop enables the bespoke agent to accommodate the full diversity of a query template's parameter space. \label{sec:mt_execution}
\revision{\paragraph{Multi-Threaded Execution as a Fifth Optimization Stage.}
Once single-threaded optimization is complete, the pipeline proceeds to a fifth stage that extends the synthesized engine to parallel execution. The agent first adds a shared \texttt{ThreadPool} that all query functions access. This pool is initialized from a \texttt{CORE\_IDS} environment variable, which determines both thread count and CPU core pinning. It then parallelizes each query independently using the same conversation-branching infrastructure as before. For each query, the agent consults the single-threaded tracing output to identify the dominant bottleneck and then selects an appropriate decomposition strategy, such as range-based scan partitioning, per-thread local aggregation with a sequential merge, or parallel hash joins. In doing so, it targets the operator that accounts for the largest share of runtime rather than parallelizing indiscriminately. Correctness and regression tracking apply throughout, exactly as in the single-threaded stages.}

%Since query templates are parameterized, different parameter instantiations can produce vastly different intermediate cardinalities and, consequently, different optimal join orders.
%Something a static cost-model would need to predict in generic DBMS implementations.
%However, cost-models for arbitrary operator implementations do not exist. 
%Hence, our bespoke-agent sidesteps cost models entirely and finds good join orders empirically: the agent proposes a join order, benchmarks it against sampled instantiations, and iterates based on observed runtime.
%The simple bootstrapping sample from \textcircled{3} prunes the most suboptimal orderings early, narrowing the search space before deeper exploration begins.
%When different parameter instantiations consistently favor different join orders, the agent has two options:
%(1) settle on a single robust ordering that performs well across the sampled instantiations, 
%or (2) hardcode separate implementations for distinct regions of the parameters space, specializing the generated code to different argument profiles.
%This try-and-evaluate loop gives the bespoke-agent the flexibility to handle the full diversity of a query template's parameter space.

\subsection{Resynthesizing with Changing Workload}
\label{subsec:changing_wl}
A bespoke OLAP engine is, by design, specialized for a fixed workload, and this naturally raises the question of what happens when the workload changes. 
Two distinct scenarios must be addressed, and each calls for a different response.

\paragraph{Ad-hoc queries outside the workload contract.}
The first scenario concerns occasional ad-hoc queries that fall outside the workload contract. 
A central invariant of the synthesis pipeline is that all data must remain materializable into a flat relational format at all times, regardless of how aggressively the bespoke engine compresses, reorders, or reorganizes it internally. 
The pipeline enforces this invariant during synthesis, which guarantees that the original relational semantics of the data are never lost, even when the internal representation deviates substantially from a conventional table layout. 
Because this invariant holds, the materialized flat representation can be handed to any general-purpose OLAP engine, such as DuckDB, which then serves as a fallback for queries that the bespoke engine was not designed to handle. 
The performance of this fallback path is naturally comparable to what the user would have obtained from a conventional system in the first place, so the worst case for an out-of-contract query is not failure but rather the absence of workload-specific optimization. 
This means that deploying a bespoke engine never makes the system less capable than a general-purpose alternative; it only adds a fast path for the queries that the contract covers.

\paragraph{Sustained workload drift.}
The second scenario concerns sustained workload drift, in which the set of queries the system is expected to execute changes substantially over time. 
In this case, the appropriate response is to resynthesize the engine entirely for the new workload. 
The key observation that makes this practical is that resynthesis is both fast and inexpensive. 
Because a full engine can be generated in the order of minutes to hours at the cost of a few dollars, resynthesis is not a catastrophic recovery procedure but can be a routine maintenance operation similar to data loading. 
This stands in sharp contrast to manually engineered specialization, where adapting to a new workload requires weeks or months of tuning effort and is therefore avoided unless absolutely necessary.

\paragraph{Modes of resynthesis.}
Furthermore, resynthesis offers two modes of operation depending on the extent of the change. 
If the new workload operates on the same schema and data but requires different query implementations or execution strategies, the engine can be resynthesized while keeping the existing storage layer intact, avoiding the cost of reingesting the data entirely and reducing deployment time. 
If the new workload also demands a fundamentally different physical layout, for instance, because new queries access different column combinations or require different sort orders, a full resynthesis including storage reorganization can be performed instead. 
In both cases, the hotpatching infrastructure described in Section~\ref{sec:system} allows the updated engine to be deployed without restarting the running process, so that the transition from the old workload to the new one happens without downtime.

\section{A System for System Generation}
\label{sec:system}

\begin{figure}[t]
\captionsetup{aboveskip=0.0ex,belowskip=0.0ex}
\captionsetup[subfigure]{aboveskip=0.0ex,belowskip=0.0ex}
    \centering
    \includegraphics[width=\linewidth]{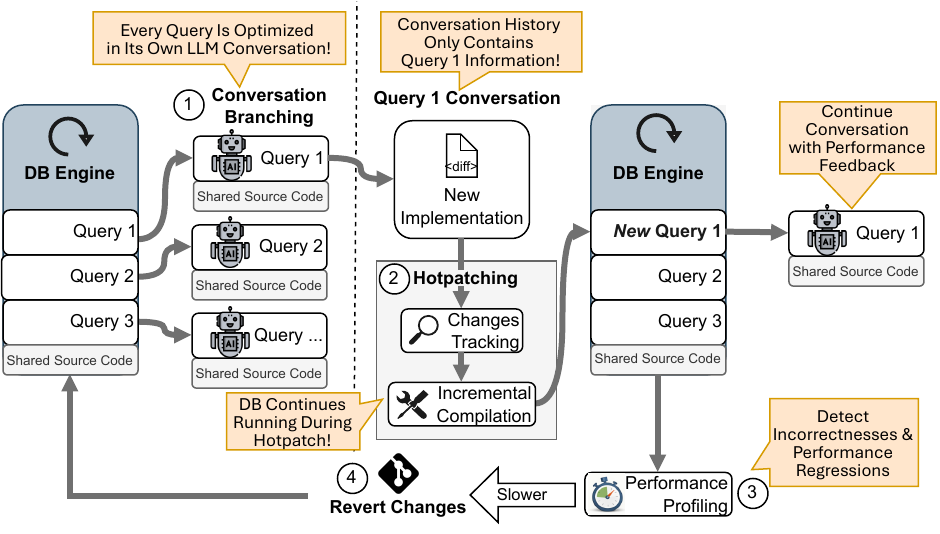}
    \Description{}
    \caption{Overview of the infrastructure supporting LLM-driven OLAP engine synthesis.
The Bespoke Agent first performs shared preparation and then branches into independent per-query optimization threads (\textcircled{1}).
Code modifications are applied through hotpatching (\textcircled{2}), allowing incremental recompilation and replacement of individual components without restarting the running database process.
Each change is automatically validated using fuzzy testing and benchmarking against DuckDB to ensure correctness and measure performance (\textcircled{3}).
All states are tracked using snapshot-based versioning with external regression monitoring (\textcircled{4}), enabling rollback, reproducibility, and monotonic optimization.
This infrastructure provides a fast, correct, and recoverable synthesis loop.}
    \label{fig:system_overview}
    %\vspace{-5ex}
\end{figure}

As shown in the previous section, synthesis of a Bespoke OLAP engine is not a single step but an incremental process that moves from a workload specification to a fully optimized engine through a sequence of well-defined stages. 
To make this process work reliably and at a reasonable cost and time, it requires more than a capable LLM. 
It requires an infrastructure that keeps the development loop fast, protects against mistakes, and ensures that every change the Bespoke Agent makes is immediately verifiable. 
This section describes both the infrastructure we built to support synthesis and the Bespoke Agent loop that uses it. 
We begin by describing the system we built for system generation, covering the mechanisms that make synthesis fast, correct, and recoverable. 

\paragraph{System Generation Requires a Dedicated Infrastructure.}
Generating a database engine with an LLM coding agent is fundamentally different from asking an LLM to write a standalone function or a small script. 
A database engine is a living system that must be compiled, loaded with data, exercised under queries, measured for correctness, and profiled for performance, and all of this must happen repeatedly and quickly during synthesis. 
Without infrastructure designed specifically to support this loop, each iteration of the Bespoke Agent's development cycle becomes prohibitively slow. 
The Bespoke Agent would need to recompile the entire codebase on every change, reload data from scratch after every modification, and wait for full system restarts before observing the effect of its edits. 
At that pace, synthesis of a complete OLAP engine would take far too long and be impractical. 
With the infrastructure we propose in the following sections, shown in \Cref{fig:system_overview} and \Cref{fig:approach_overview}, synthesis can be completed in minutes to hours rather than days.

\paragraph{Conversation Branching to Enable Per-Query Optimization.} \label{subsec:ConversationBranching}
% A better insight into synthesizing a system for a workload is that managing the Bespoke Agent's conversation history, the context used as input to the Bespoke Agent, throughout incremental generation.
Another insight into synthesizing a system for a workload is the importance of managing the Bespoke Agent's conversation history (the context the agent operates within) throughout incremental generation.
Overall, managing this context is as important as managing the code it produces. 
As discussed before, during the system generation phase, at the beginning, some steps are shared across all queries. 
We therefore begin the generation with a shared conversation for all queries in which the Bespoke Agent performs global preparation, including planning storage and generating the storage code. 
Inconsistency in the storage layer, the query interface, and how pinning and tracing components are used across queries would prevent any deep optimizations, and make the output result unreliable and performance unstable. 
\revision{The shared foundation consists of three concrete artifacts: an implementation plan specifying the storage approach and data structures, the builder code (\texttt{builder.cpp}) that ingests Arrow/Parquet tables into the in-memory layout, and query interface stubs with agreed-upon function signatures. The foundation is complete once the builder compiles and the ingest path is validated against a reference implementation.}
Once the shared foundation is established, however, we break the conversation into a separate thread per query (called conversation branching). This can be seen at step \textcircled{1} of Figure \ref{fig:system_overview} where we split the context per query.

For the branching conversation, we split the synthesis into multiple Bespoke Agents, one agent thread per query template.
Each per-query agent thread carries the shared history as its starting point, but then develops its context independently, focusing entirely on optimizing a single query without being distracted by the implementation details of others. 
This is possible because queries can be optimized independently without risk of interference. 
However, the shared components, such as storage, can still be updated by individual agents to improve storage based on insights gained when synthesizing code for a single query.
To enable this and still have one shared storage for all queries, each agent contributes their changes to the shared sequence so that changes to shared components remain consistent. 
Concretely, the first per-query Bespoke Agent commits its changes before the second begins, and so on through the sequence. 
Importantly, the order in which queries are processed does not affect correctness or performance, as multiple rounds of implementations, evaluations and optimizations are applied.
Because each branch operates live on the same running database engine, each per-query Bespoke Agent can observe any changes by a preceding Bespoke Agent, including both general changes on the engine and interfaces, as well as the storage layout, and can therefore build correctly on top of them. 
The result is that each Bespoke Agent reasons within a compact and focused context while still operating on a globally consistent codebase.

% \paragraph{Hotpatching Keeps the Engine Running While Code Evolves.}
\paragraph{Fast Iteration through Continuous and Restart-Free Deployment.}
The most consequential infrastructure component is the hotpatching engine, which can be seen in step \textcircled{2} of Figure \ref{fig:system_overview}. 
Rather than treating the synthesized database as a static binary that must be fully recompiled and restarted after every change, our system keeps the database process running continuously throughout synthesis. 
When the Bespoke Agent modifies a specific component of the engine, such as the execution function for a single query or a helper data structure, the system recompiles only that component in isolation and swaps it into the running process without interruption. 
The consequence of this design is significant. 
Because the process never stops, the storage layer remains resident in memory at all times. 
This means the Bespoke Agent never needs to re-run the ingestion pipeline after a code change, which at a large scale factor of data would otherwise consume minutes per iteration. 
\change{Only if the Bespoke Agent decides to change the storage layer is a reload of the data necessary.}{A data reload is only necessary when the Bespoke Agent decides to modify the storage layer.}
Combined with incremental recompilation, hotpatching reduces the turnaround time between a Bespoke Agent edit and an observable result from minutes to seconds. 
This reduction is not merely a convenience. 
With hundreds of patching steps required to produce a full, highly optimized TPC-H engine, this expansion increases the number of optimization candidates the Bespoke Agent can explore within a fixed time budget and therefore affects the quality of the final synthesized engine.

%\paragraph{Planning the storage to enable interaction-minimal query implementation.} - first plan the storage - implement ingest/build - add a naive implementation of the queries to test the vality of our ingest implementation / correctness of the loaded data / data at rest. \todo{new paragraph here?}

\paragraph{Correctness needs Continuous \& Fast Testing.}
Correctness validation in our pipeline (see \textcircled{3} in Figure \ref{fig:system_overview}) goes beyond checking that a query returns the right result for a single fixed input. 
Because query templates are parameterized, a query implementation may be correct for one parameter instantiation and subtly wrong for another. 
To catch such cases, the benchmark tool for the Bespoke Agent runs each query template across a diverse set of parameter instantiations within the allowed ranges defined in the workload specification. 
This testing approach significantly increases the probability of detecting implementation errors that a single fixed test case would miss. 
The results are compared against DuckDB for every instantiation.
To speed up validation time and ensure that correctness checks are cheap, validation runs on a smaller scale factor during the implementation phase to keep turnaround fast.
For this, using the dataset scaler of \cite{graceful}, testing is repeated on several scaled-down versions of the dataset to ensure consistency with slightly different data, preventing hardcoding to exact row-Ids or counts.
We only switch to the full dataset size during the performance optimization phase, where runtime measurements must reflect production conditions.
However, if computational cost and synthesis time are a preference, the full pipeline could run on scaled-down versions of the datasets and extrapolate performance improvements to the full dataset.

\paragraph{Importance of Regression Tracking \& Rollbacks.}
LLM coding agents are effective at generating code changes but are systematically unreliable at recognizing when a change has made things worse. 
A Bespoke Agent that has just introduced a regression will often attempt to fix it with further changes, compounding the problem rather than reverting it. 
To prevent this, our system tracks the performance impact of every change the Bespoke Agent makes through an external monitor that is entirely outside the Bespoke Agent's control. 
After each change, the benchmark tool reports the updated runtime for all affected queries, and the external monitor compares this against the most recent correct code base, which we call the \emph{accepted baseline}. 
If a regression is detected, the monitor immediately rolls back the Bespoke Agent's changes to the last known-good state and signals the Bespoke Agent to explore a different direction. 
% Because rollback is handled externally outside the Bespoke Agent, the Bespoke Agent cannot accidentally persist a regressed state if it fails to recognize the degradation itself. 
This ensures that in case the Bespoke-Agent fails to recognize its own degradation, it cannot accidentally continue on a regressed state and potentially get stuck in a suboptimal direction.
This mechanism ensures that the synthesis process is monotonically improving and that the Bespoke Agent's exploration of the optimization space never undoes progress that has already been validated.

\paragraph{Code Versioning Enables Reproducibility and Recovery.}
Every incremental change the Bespoke Agent makes to the codebase is tracked by a high-performance snapshotting system that we have built on top of Git (see \textcircled{4} of Figure \ref{fig:system_overview}). 
This versioning layer serves three purposes. 
First, it is the foundation on which the rollback process operates, since every snapshot represents a state to which the monitor can return if a regression is detected. 
Second, it enables full reproducibility of the synthesis process, because the complete sequence of code states from initial storage layout through final optimized engine can be replayed at any point. 
This is particularly valuable when a human engineer wishes to inspect or modify a specific intermediate state or to adjust a prompt and observe the downstream effect without re-running the entire synthesis from scratch. 
Third, versioning provides a durable backup of all intermediate states, which protects against infrastructure failures that would otherwise require synthesis to restart from the beginning.
Together, these properties make the synthesis process not just fast and reliable but also transparent and recoverable at every stage.
\revision{\paragraph{Lessons Learned: Infrastructure as the Enabler of LLM-Driven Synthesis.}\label{subsec:lessonsLearned}
Taken together, these infrastructure components reflect a central lesson from building Bespoke OLAP: LLMs perform substantially better when presented with specific and bounded subproblems. Hotpatching, conversation branching, context compaction, fuzzy correctness testing, and external regression monitoring all exist for this reason. Each component reduces the complexity of the problem the agent faces at any given moment, ensuring it always operates on a concrete and verifiable task rather than an underspecified global objective. The value of the pipeline is therefore not in making the LLM smarter, but in structuring the synthesis task so that the LLM's strengths are exploited and its failure modes are contained.}
\section{Experimental Evaluation}
\label{sec:evaluation}
In this section, we evaluate the impact of synthesizing bespoke OLAP engines. We demonstrate our approach on two established workloads from benchmarks:  TPC-H and CEB, and organize our evaluation around three central questions: (1)~What speedups does Bespoke-OLAP achieve over a state-of-the-art general-purpose OLAP engine? (2)~Where do these speedups originate - what are the individual contributions of the optimization stages, the bespoke storage design, and the specific strategies the agent employs? (3)~How does the synthesis process itself behave in terms of development effort, tool usage, and correctness?

\subsection{Experiment Setup}
\label{subsec:setup}
\paragraph{Benchmarking Workloads \& Hardware.}
We employ two complementary benchmarks to evaluate Bespoke-OLAP. TPC-H is a leading industrial benchmark for OLAP systems that features well-defined analytical queries for which existing engines are heavily optimized. To complement TPC-H, we additionally chose CEB~\cite{flowlossceb}, a templated version of the JOB benchmark~\cite{leis2015good} that operates on the real-world IMDB dataset. For both benchmarks, we follow established data generation and instantiation procedures to ensure reproducibility. TPC-H data was generated according to the TPC-H v3.0.1 specification~\cite{tpch} using the DuckDB data generator. For CEB, we sampled query instantiations from the published artifacts of~\cite{flowlossceb} and used the data downscaler and upscaler provided by~\cite{graceful} to obtain different scale factors. 
All experiments were executed on an AMD EPYC 9654P 96-Core Processor with 768\,GB RAM. 

\paragraph{Bespoke Engines and Baseline.}
We ran our Bespoke-OLAP agent on both benchmarks separately, producing two distinct bespoke engines: Bespoke-TPCH and Bespoke-CEB. Both implementations were created using the same framework with identical tools, prompts, and configurations. No manual interventions were performed at any stage.
As a baseline, we use DuckDB (version 1.4.1), a state-of-the-art embedded analytical database engine. \revision{We additionally compare against Umbra (version 26.02), a state-of-the-art compiled-query engine that already eliminates query-interpretation overhead.} To ensure a controlled comparison, DuckDB, Bespoke-OLAP and Umbra operate in single-threaded mode, pinned to the same core, in all experiments except the multi-threaded scaling study in Section~\ref{sec:MT-scaling}. Furthermore, to guarantee all data resides in main memory, and DuckDB/Umbra has no I/O disadvantage, we loaded data from Parquet files into DuckDB/Umbra tables within an in-memory session before each experiment.

\paragraph{LLM Models and Configuration.}\label{par:LLMModelsAndConfig}
\change{We use GPT5.2 Codex (latest at submission)\cite{openai_codex2025} due to its strong performance on large codebases and reliable tool use. Compared to GPT5 and GPT5.1 Codex, it yields fewer coding errors and more stable behavior under long contexts. As GPT5.2 Codex is trained for the Shell and Patch tools in the Agents SDK, and our experiments confirm clear benefits, we adopt the Agents SDK as execution backend; custom agents led to higher token usage and reduced stability. GPT5.2 Codex further provides roughly $2\times$ lower per-token cost than Opus4.6 and supports controllable context compaction, enabling task-aware optimization. For these reasons (native SDK training, lower cost, and controllable compaction), we focus exclusively on Codex models.}{All experiments in this paper use Claude Sonnet 4.7 as the primary model. We selected it for its speed, cost-efficiency, and reliable tool use across long synthesis runs. However, the pipeline is model-agnostic by design, and we have tested it with a range of other models. Claude Sonnet~4.7 and GPT~5.4-Codex achieved comparable speedups. We additionally tested locally-hosted open-weight models, including Gemma-4-26B-A4B-it, MiniMax-M2.7, GLM-4.7-FP8, and GLM-5.1-FP8, which generally achieved lower speedups and required more attempts per stage. We attribute this to model capability and instruction-following quality rather than a fundamental limitation of the approach.}
\begin{comment}
We use GPT~5.2 Codex (the most recent version at the time of submission) as the underlying language model. We chose Codex because it is a highly capable coding model that handles large codebases with very high accuracy. Compared to its predecessors (GPT~5 Codex and GPT~5.1 Codex), we observed significantly fewer coding errors and more reliable tool use, particularly as the context grew larger. OpenAI states that GPT~5.2 Codex was specifically trained for use with the Shell and Patch tools provided through the OpenAI Agents SDK. After initial experiments confirmed this advantage, we adopted the Agents SDK as our execution backend. A self-implemented agent handling resulted in drastically higher token usage, as the model struggled with a custom setup it was not trained on. Additionally, the per-token cost of GPT~5.2 Codex is roughly $2\times$ lower compared to Anthropic's frontier coding model (Opus~4.6), and OpenAI's SDK provides easily controllable compaction - allowing us to leverage our domain knowledge about the task structure and stage ordering to determine optimal compaction points. For these three reasons - native Agents SDK training, a lower price point, and easily controllable compaction - we focus exclusively on Codex models. Nonetheless, we believe the same approach, with appropriate modifications to the tool interfaces and compaction strategy, will generalize to other coding agents as well.
\end{comment}

\subsection{Performance and Scalability}
\label{subsec:perf_scalability}
%We begin by examining the end-to-end performance. %: the overall speedups relative to DuckDB, per-query behavior, and scalability across different dataset sizes.

\begin{figure*}[t]
\captionsetup{aboveskip=0.0ex,belowskip=0.0ex}
\captionsetup[subfigure]{aboveskip=0.0ex,belowskip=0.0ex}
    \centering
    \revisionimage{
    \includegraphics[width=\linewidth]{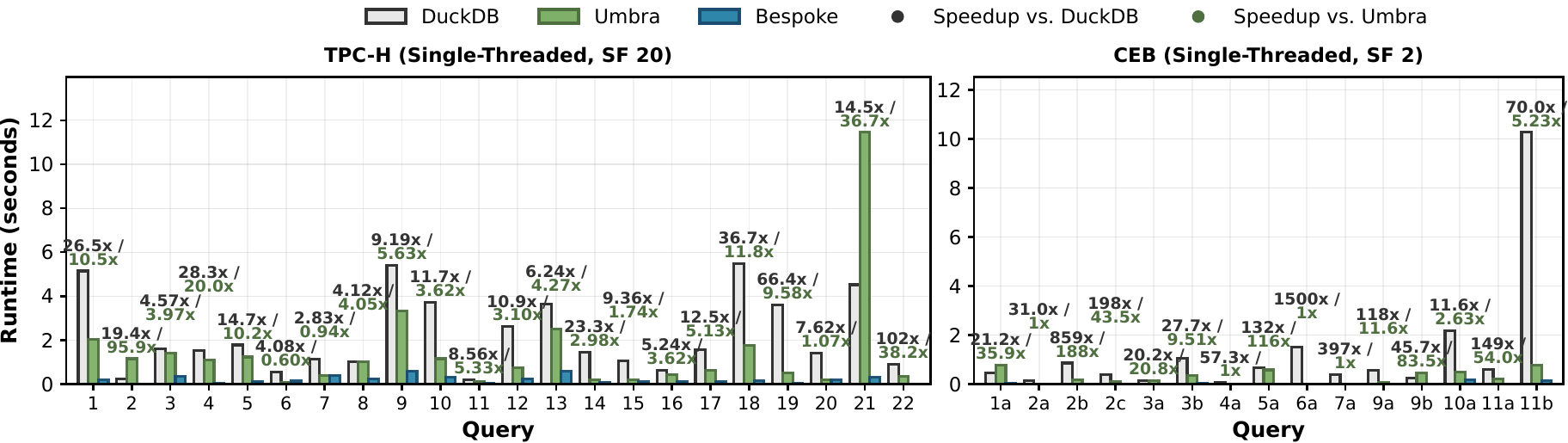}
    }
    \vspace{-1ex}
    \Description{Per-query  runtimes comparing Bespoke-TPCH/CEB and \revision{Umbra} on TPC-H (SF20) and CEB (SF2) \revision{single-threaded}.}
    \caption{Per-query  runtimes comparing Bespoke-TPCH/CEB and \revision{Umbra} on TPC-H (SF20) and CEB (SF2) \revision{single-threaded}.}
    \label{fig:per_query_tpch}
    \vspace{-2ex}
\end{figure*}

\subsubsection{Overall and Per-Query Speedups}
\label{subssubsec:OverallSpeedups}
Figure \ref{fig:speedup} presents the overall speedup achieved by Bespoke-OLAP. On TPC-H at scale factor~20, Bespoke-TPCH completes the full workload in 4.4\,s compared to 49.2\,s for DuckDB, yielding an overall speedup of \SpeedupSTDuckTPCH{}. On CEB at scale factor~2, Bespoke-CEB finishes in 0.4\,s versus 19.5\,s for DuckDB, corresponding to a speedup of \SpeedupSTDuckCEB{}.\footnote{Query~8a of CEB has been excluded from all plots. DuckDB exhibited vastly spiking runtimes depending on the query instantiation, ranging from 10$\times$ to over 100$\times$. Due to these highly unstable results, Q8a was excluded. It is worth noting, however, that Bespoke-CEB maintains very stable runtimes across all instantiations of this query.}
Crucially, these speedups are not driven by a few outlier queries. As Figure \ref{fig:per_query_tpch} reveals, Bespoke-OLAP outperforms DuckDB on \emph{every single query}, with per-query speedups ranging from 2.83$\times$ to 102$\times$ on TPC-H and from 11.6$\times$ to 1500$\times$ on CEB. The consistently high speedups across all queries underscore that Bespoke-OLAP's advantages are systematic rather than accidental.
\revision{In comparison to Umbra, Bespoke-OLAP achieves overall speedups of \SpeedupSTUmbraTPCH{} on TPC-H and \SpeedupSTUmbraCEB{} on CEB over Umbra (single threaded execution). Remarkably, Bespoke-OLAP outperforms Umbra on all but two queries across both benchmarks combined, with the two exceptions at 0.60$\times$ and 0.94$\times$. This demonstrates that the generality tax extends beyond query compilation into storage layout and algorithmic design: even a compiled-query engine that eliminates interpretation overhead cannot match a workload-specific engine that also optimizes storage organization and query algorithms for the given workload.}
The high-speedup gains on TPC-H also demonstrate the overall efficiency of Bespoke engines, as OLAP engines are already highly tuned for TPC-H and sometimes even implement specific code paths to execute such queries optimally, achieving good results on the TPC-H benchmark leaderboard.

% \begin{figure*}
%     \centering
%     \includegraphics[width=.8\linewidth]{figures/journal_per_query_absolute_runtime_tpch.pdf}
%     \Description{Per-query absolute runtimes comparing Bespoke-TPCH and DuckDB on TPC-H SF20.}
%     \caption{Per-query absolute runtimes for TPC-H at SF20. Bespoke-TPCH outperforms DuckDB on every query.}
%     \label{fig:per_query_tpch}
% \end{figure*}

% \begin{figure*}
%     \centering
%     \includegraphics[width=.8\linewidth]{figures/journal_per_query_absolute_runtime_ceb.pdf}
%     \Description{Per-query absolute runtimes comparing Bespoke-CEB and DuckDB on CEB SF2.}
%     \caption{Per-query absolute runtimes for CEB at SF2\protect\footnotemark}
%     \label{fig:per_query_ceb}
% \end{figure*}

\begin{figure}
\captionsetup{aboveskip=0.0ex,belowskip=0.0ex}
\captionsetup[subfigure]{aboveskip=0.0ex,belowskip=0.0ex}
    \centering
    \revisionimage{
    \includegraphics[width=.9\linewidth]{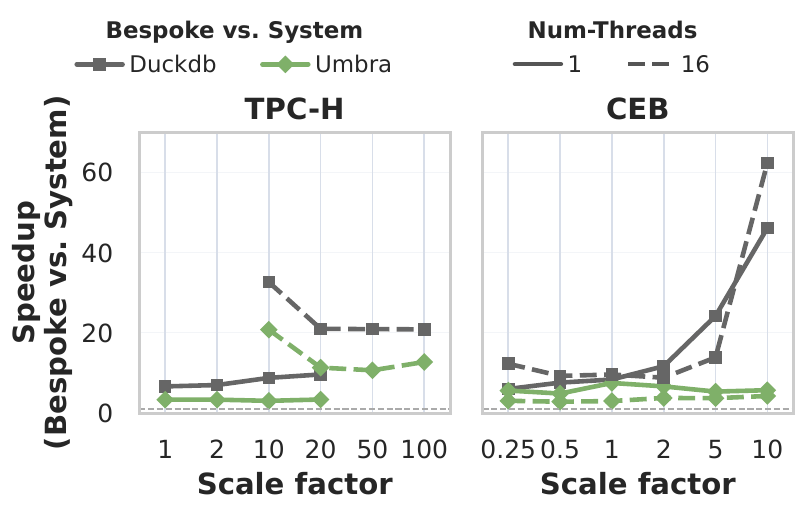}
    }
    \Description{Speedup of Bespoke-OLAP over DuckDB at different scale factors for TPC-H and CEB.}
    \caption{Speedups over DuckDB and Umbra at different scale factors for TPC-H (left) and CEB (right). On TPC-H, the speedup remains stable. On CEB, Bespoke-OLAP scales increasingly better than DuckDB, with the speedup growing to over 45$\times$ at the largest scale factor. This stems from an algorithmically lower complexity class in Bespoke-CEB compared to DuckDB, producing increasing speedups with larger data sizes.}
    \label{fig:scale_factor_comparison}
    \vspace{-2ex}
\end{figure}

\subsubsection{Scalability with Increasing Data Sizes}
We now examine if the performance advantages persist as dataset sizes grow.
Figure \ref{fig:scale_factor_comparison} shows that the synthesized engines scale gracefully. On TPC-H, the speedup remains stable across all tested scale factors, indicating that Bespoke-TPCH and DuckDB exhibit the same asymptotic complexity on this workload. On CEB, however, Bespoke-OLAP scales \emph{significantly better} than DuckDB: the speedup grows substantially with increasing scale factors, reaching over 45$\times$ at scale factor~10, as DuckDB's runtime grows disproportionately. This widening gap on CEB can be attributed to Bespoke-CEB's carefully tuned storage layouts and query implementations, which are more robust to the increased data volumes and the complex, correlated data distributions present in the IMDB dataset. Together, these results demonstrate that Bespoke-OLAP not only delivers substantial speedups at a fixed scale, but that these advantages are maintained - and in some cases amplified - as the data grows.

\begin{figure}
\captionsetup{aboveskip=0.0ex,belowskip=0.0ex}
    \centering
    \revisionimage{
    \includegraphics[width=\linewidth]{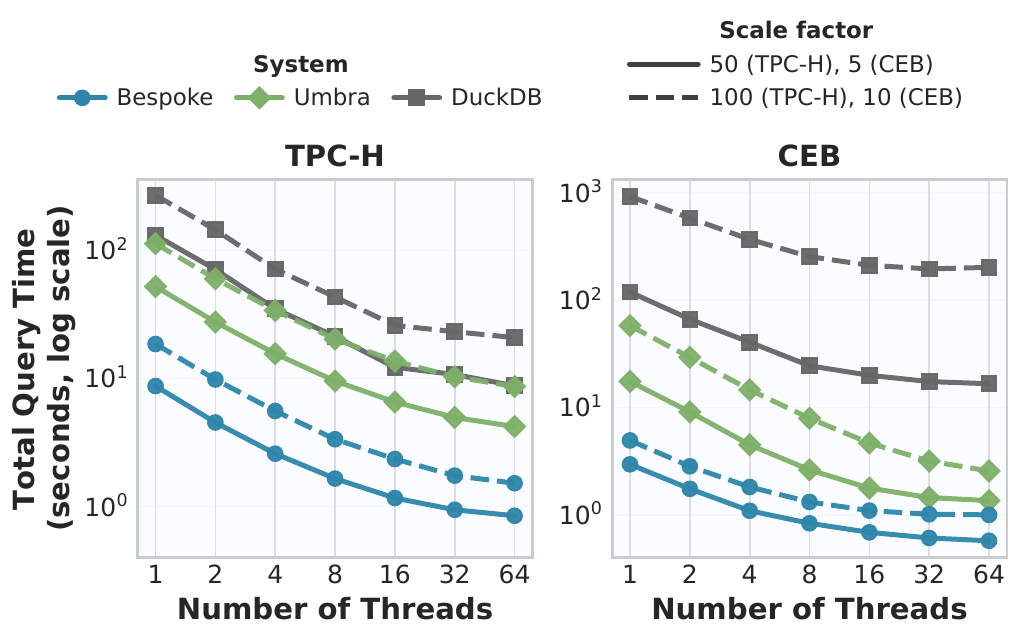}
    }
    \vspace{-2ex}
    \caption{Multi-threaded scaling on TPC-H and CEB.
    Bespoke remains the fastest system at every thread count.
    In multi-threaded execution, Bespoke achieves \SpeedupMTDuckTPCH{}
    over DuckDB and \SpeedupMTUmbraTPCH{} over Umbra on TPC-H.
    On CEB the speedups are \SpeedupMTDuckCEB{} over DuckDB
    and \SpeedupMTUmbraCEB{} over Umbra.}
    \label{fig:mt_scaling}
    \vspace{-2ex}
\end{figure}

\subsubsection{Scalability with Increasing Number of Threads} \label{sec:MT-scaling}
Figure~\ref{fig:mt_scaling} shows total query time as the number of threads increases from 1 to 64 on both benchmarks. Bespoke remains the fastest system at every thread count on both benchmarks. In multi-threaded execution, Bespoke achieves \SpeedupMTDuckTPCH{} over DuckDB and \SpeedupMTUmbraTPCH{} over Umbra on TPC-H. On CEB the speedups are \SpeedupMTDuckCEB{} over DuckDB and \SpeedupMTUmbraCEB{} over Umbra.

\begin{figure}
\captionsetup{aboveskip=0.0ex,belowskip=0.0ex}
\captionsetup[subfigure]{aboveskip=0.0ex,belowskip=0.0ex}
    \centering
    \revisionimage{
    \includegraphics[width=\linewidth]{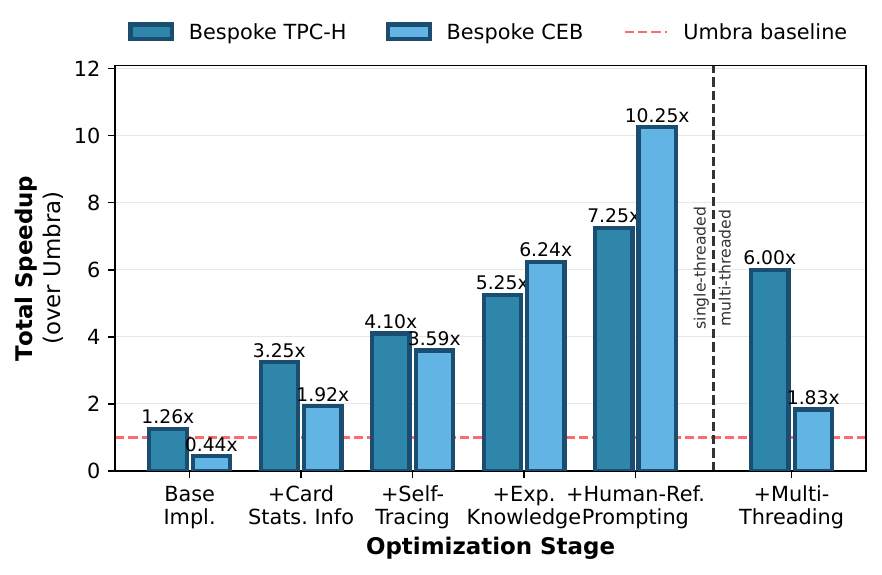}
    }
    \vspace{-2.5ex}
    \Description{Ablation study showing the cumulative impact of each optimization stage on overall speedup for TPC-H and CEB.}
    \caption{\revision{}Cumulative impact of each optimization stage on the overall speedup for TPC-H and CEB. Each bar includes all optimizations from preceding stages.}
    \label{fig:optim_ablation}
    \vspace{-3ex}
\end{figure}

\subsection{Anatomy of the Speedups}
\label{subsec:anatomy}
The previous experiment established \emph{what} Bespoke-OLAP achieves; we now turn to understanding \emph{where} these speedups originate. We decompose the performance gains along three dimensions: the incremental impact of each optimization stage, the contribution of bespoke storage layouts versus query-execution code, and the concrete strategies the agent employs at the code level.

\subsubsection{Impact of Optimization Stages}\label{subsec:ImpactOptiStages}
As described in \ref{sec:approach}, the pipeline executes four consecutive optimization stages. Figure \ref{fig:optim_ablation} shows the measured speedup after each stage, where each stage includes all optimizations from preceding stages.
Importantly, we can see that every optimization stage contributes meaningfully, confirming the value of the multi-stage pipeline design.

On TPC-H, the initial implementation already achieves a speedup of 2.34$\times$ over DuckDB, even though the generated query-execution code is entirely unoptimized. This initial gain stems solely from the bespoke storage design, which avoids the overhead of a general-purpose storage layer such as schema interpretation. \revision{To give some examples of what drives this: \texttt{lineitem} is sorted primarily by \texttt{l\_shipdate} and secondarily by \texttt{l\_orderkey}, enabling zone-map pruning that skips entire  64\,K-row blocks for the date-range filters present in fourteen of the twenty-two  TPC-H queries. The four most frequently aggregated columns are packed into an 8-byte AoS struct so that a single cache-line fetch retrieves all four values per row. Furthermore, a  pre-computed derived column stores the discounted price, eliminating a per-row multiply in the innermost loop of thirteen queries. The full set of storage decisions is documented in this artifact.\footnote{The storage plan can be found \href{https://github.com/DataManagementLab/BespokeOLAP_Artifacts/blob/main/bespoke_tpch_multithreading/storage_plan.txt}{here}.}}

On CEB, by contrast, the initial implementation achieves only 2.10$\times$ over DuckDB and remains at a \emph{slowdown} relative to Umbra (0.44$\times$). Here, the more complex access patterns arising from the real-world IMDB dataset, with its irregular data distributions and many-to-many relationships, require more advanced processing strategies that the initial execution code cannot yet provide; the bespoke storage alone is insufficient to compensate. Subsequent optimization stages progressively improve performance. With additional actual cardinality information, self-tracing of the generated implementation, and the injection of expert knowledge, each stage contributes measurable speedups on both benchmarks. Notably, the final stage, in which we provided the agent with a human reference persona for the code it was writing and the optimizations it was performing, yielded a substantial additional improvement. \revision{To give a concrete example: for Query~1, the prior stages had produced a sequential scan of \texttt{l\_shipdate} with 4-way loop unrolling and branch-prediction hints. The human persona stage instead restructured the storage builder to partition the \texttt{lineitem} rows into nine groups by (returnflag, linestatus) key, sorting each group by shipdate at ingest time. This allowed the query to replace the full scan with a binary search per group, followed by a pure sequential AVX-512 reduction processing 16 elements per cycle with two independent accumulator vectors for instruction-level parallelism. The result is near-peak memory bandwidth utilization. The final Bespoke-TPC-H engine is approximately 11,600 lines of C++.} This step led the model to exploit techniques it had not previously applied, such as more aggressive low-level optimizations and tighter data-access patterns, resulting in final speedups of 12.35$\times$ on TPC-H and 51.40$\times$ on CEB over DuckDB. 

\subsubsection{Impact of Bespoke Storage}
The optimization ablation demonstrates that each stage contributes, but it does not disentangle the effects of bespoke \emph{storage} from bespoke \emph{query execution}. To isolate these two dimensions, we ran the Bespoke-OLAP agent when fixing storage to a flat design, a simple struct-of-arrays representation with no workload-specific tuning and compared it to the full bespoke generation with bespoke storage.

\begin{table}[t]
\small
{
    \scalebox{1.0}{
    \begin{tabular}{c|rr|rr}
    \hline
     & \multicolumn{2}{c|}{\textbf{w/ Flat Storage}} & \multicolumn{2}{c}{\textbf{w/ Bespoke Storage}} \\
     & \multicolumn{1}{c}{\begin{tabular}[c]{@{}c@{}}Basic \\ Impl.\end{tabular}} & \multicolumn{1}{c|}{\begin{tabular}[c]{@{}c@{}}After Optim.\\ Loop\end{tabular}} & \multicolumn{1}{c}{\begin{tabular}[c]{@{}c@{}}Basic \\ Impl.\end{tabular}} & \multicolumn{1}{c}{\begin{tabular}[c]{@{}c@{}}After Optim.\\ Loop\end{tabular}} \\ \hline
    \textbf{TPC-H} & 1.26  & 5.18 & 2.34 & 12.35 \\
    \textbf{CEB} & 0.57 & 8.09 & 2.10 & 51.40 \\ \hline
    \end{tabular}}}
    \caption{Speedups over DuckDB with flat storage (struct-of-arrays) vs. bespoke storage. Flat storage shows the benefits of bespoke query code only; bespoke storage adds workload-specific physical data layouts.}
    \label{tab:bespoke_storage_impact}
    \vspace{-7ex}
\end{table}

Table \ref{tab:bespoke_storage_impact} shows the results, which reveal two important findings. First, even with a flat storage layout, the Bespoke-OLAP agent achieves speedups of 1.26$\times$ on TPC-H and approaches parity on CEB (0.57$\times$) in the basic implementation phase alone. These gains, modest as they are, arise purely from removing the layers of abstraction inherent in a general-purpose query engine: the generic processing model, the iterator framework, and the type-dispatching overhead. This confirms that \emph{the agent can generate efficient query-processing code} independent of the storage design. Second, when enabling bespoke storage, the speedups increase dramatically to 12.35$\times$ on TPC-H and 51.40$\times$ on CEB. The bespoke storage component thus amplifies the gains from efficient query execution by providing physical data layouts tailored to each query's access patterns. Taken together, these results demonstrate that Bespoke-OLAP derives its performance from a synergy of two capabilities: generating efficient, abstraction-free query-processing code \emph{and} designing highly beneficial storage layouts.

\subsubsection{Analysis of Employed Strategies}
To gain deeper insight into \emph{how} the Bespoke Agent achieves these speedups, we systematically classified the strategies employed across the storage, query-execution, and low-level optimization layers in both workloads.

\definecolor{grp40636442}{rgb}{0.941,0.992,0.957}
\definecolor{grp90418026}{rgb}{0.996,0.949,0.949}
\definecolor{grp65316561}{rgb}{0.925,0.996,1.000}
\definecolor{grp14611914}{rgb}{1.000,0.969,0.929}
\definecolor{grp11059936}{rgb}{0.941,0.976,1.000}
\definecolor{grp25827640}{rgb}{0.980,0.961,1.000}
\begin{table}[t]
\scalebox{0.83}{
% ----- Auto generated by `classify_data.ipynb` -----
% ----- Copy into your LaTeX document -----
\begin{tabular}{l l c r}
\hline
\small
 & \shortstack[c]{Bespoke and Non-Bespoke\\Database Strategies} & \shortstack[c]{Classical\\/Exotic} & used in \% of \\
\hline
\cellcolor{grp40636442}  & \cellcolor{grp40636442} Dictionary Encoding & \cellcolor{grp40636442} C & \cellcolor{grp40636442} 25.00\%/cols \\
\cellcolor{grp40636442}  & \cellcolor{grp40636442} Physical Row Sort Order & \cellcolor{grp40636442} C & \cellcolor{grp40636442} 18.75\%/cols \\
\cellcolor{grp40636442}  & \cellcolor{grp40636442} Flat String Arena (offset array) & \cellcolor{grp40636442} C & \cellcolor{grp40636442} 10.94\%/cols \\
\cellcolor{grp40636442}  & \cellcolor{grp40636442} Scaled Integer / Fixed-Point & \cellcolor{grp40636442} E & \cellcolor{grp40636442} 7.03\%/cols \\
\cellcolor{grp40636442}\multirow{-5}{*}{\rotatebox[origin=c]{90}{\shortstack{Column\\Encoding\\Strategies}}} & \cellcolor{grp40636442} Compact Date (16-bit offset) & \cellcolor{grp40636442} E & \cellcolor{grp40636442} 1.56\%/cols \\
\hline
\cellcolor{grp90418026}  & \cellcolor{grp90418026} Hash Index (O(1) join lookup) & \cellcolor{grp90418026} E & \cellcolor{grp90418026} 47.37\%/queries \\
\cellcolor{grp90418026}  & \cellcolor{grp90418026} Precomputed Derived Column & \cellcolor{grp90418026} E & \cellcolor{grp90418026} 36.84\%/queries \\
\cellcolor{grp90418026}  & \cellcolor{grp90418026} Sorted Range Directory (join key) & \cellcolor{grp90418026} C & \cellcolor{grp90418026} 26.32\%/queries \\
\cellcolor{grp90418026}\multirow{-4}{*}{\rotatebox[origin=c]{90}{\shortstack{Query-\\Support\\Structures}}} & \cellcolor{grp90418026} \shortstack[l]{Precomp. Arith. Lookup Table} & \cellcolor{grp90418026} E & \cellcolor{grp90418026} 2.63\%/queries \\
\hline
\cellcolor{grp65316561}  & \cellcolor{grp65316561} Dictionary Predicate Rewrite & \cellcolor{grp65316561} E & \cellcolor{grp65316561} 71.05\%/queries \\
\cellcolor{grp65316561}  & \cellcolor{grp65316561} Sorted Key Range Lookup & \cellcolor{grp65316561} C & \cellcolor{grp65316561} 18.42\%/queries \\
\cellcolor{grp65316561}  & \cellcolor{grp65316561} Shard-Skip Scan (Zone Maps) & \cellcolor{grp65316561} C & \cellcolor{grp65316561} 13.16\%/queries \\
\cellcolor{grp65316561}\multirow{-4}{*}{\rotatebox[origin=c]{90}{\shortstack{Scan\\\& Filter\\Strategies}}} & \cellcolor{grp65316561} Sorted Range Scan & \cellcolor{grp65316561} C & \cellcolor{grp65316561} 13.16\%/queries \\\hline
\cellcolor{grp14611914}  & \cellcolor{grp14611914} Bitmap Semi-Join & \cellcolor{grp14611914} E & \cellcolor{grp14611914} 73.68\%/queries \\
\cellcolor{grp14611914}  & \cellcolor{grp14611914} Index Nested-Loop Join & \cellcolor{grp14611914} C & \cellcolor{grp14611914} 42.11\%/queries \\
\cellcolor{grp14611914}  & \cellcolor{grp14611914} Tag-Array Join & \cellcolor{grp14611914} E & \cellcolor{grp14611914} 26.32\%/queries \\
\cellcolor{grp14611914}  & \cellcolor{grp14611914} Hash Join & \cellcolor{grp14611914} C & \cellcolor{grp14611914} 10.53\%/queries \\
\cellcolor{grp14611914}\multirow{-5}{*}{\rotatebox[origin=c]{90}{\shortstack{Join\\Operators}}} & \cellcolor{grp14611914} Sort-Merge Join & \cellcolor{grp14611914} C & \cellcolor{grp14611914} 2.63\%/queries \\
\hline
\cellcolor{grp11059936}  & \cellcolor{grp11059936} Inline (Fused) Aggregation & \cellcolor{grp11059936} E & \cellcolor{grp11059936} 97.37\%/queries \\
\cellcolor{grp11059936}  & \cellcolor{grp11059936} Dense-Key Aggregation & \cellcolor{grp11059936} E & \cellcolor{grp11059936} 55.26\%/queries \\
\cellcolor{grp11059936}  & \cellcolor{grp11059936} Scalar Aggregation & \cellcolor{grp11059936} C & \cellcolor{grp11059936} 39.47\%/queries \\
\cellcolor{grp11059936}  & \cellcolor{grp11059936} Two-Phase Aggregation & \cellcolor{grp11059936} C & \cellcolor{grp11059936} 39.47\%/queries \\
\cellcolor{grp11059936}  & \cellcolor{grp11059936} Direct Array Aggregation & \cellcolor{grp11059936} E & \cellcolor{grp11059936} 28.95\%/queries \\
\cellcolor{grp11059936}\multirow{-6}{*}{\rotatebox[origin=c]{90}{\shortstack{Aggregation\\Strategies}}} & \cellcolor{grp11059936} Hash Aggregation & \cellcolor{grp11059936} C & \cellcolor{grp11059936} 18.42\%/queries \\
\hline
\cellcolor{grp25827640}  & \cellcolor{grp25827640} Pointer \_\_restrict Aliasing & \cellcolor{grp25827640} E & \cellcolor{grp25827640} 50.00\%/queries \\
\cellcolor{grp25827640}  & \cellcolor{grp25827640} Integer Scaled Arithmetic & \cellcolor{grp25827640} E & \cellcolor{grp25827640} 44.74\%/queries \\
\cellcolor{grp25827640}  & \cellcolor{grp25827640} Branchless Range Check & \cellcolor{grp25827640} E & \cellcolor{grp25827640} 18.42\%/queries \\
\cellcolor{grp25827640}  & \cellcolor{grp25827640} Branch-Hint (\_\_builtin\_expect) & \cellcolor{grp25827640} E & \cellcolor{grp25827640} 13.16\%/queries \\
\cellcolor{grp25827640}  & \cellcolor{grp25827640} Manual Loop Unrolling & \cellcolor{grp25827640} E & \cellcolor{grp25827640} 13.16\%/queries \\
\cellcolor{grp25827640}\multirow{-6}{*}{\rotatebox[origin=c]{90}{\shortstack{Low-Level\\Execution\\Optimizations}}} & \cellcolor{grp25827640} Software Prefetching & \cellcolor{grp25827640} E & \cellcolor{grp25827640} 10.53\%/queries \\
\hline
\end{tabular}
% -- ENd auto generated
}

\caption{Overview of bespoke strategies employed in the storage layout and query-processing implementations across Bespoke-TPCH and Bespoke-CEB. We classify the strategies into classical (C) and exotic (E). Classical ones can also be found in general-purpose OLAP engines, but here they are used in a workload-specific manner. Exotic ones are strategies that are not used in relational engines.}
\label{tab:code_classification}
\vspace{-7.5ex}
\end{table}

Table \ref{tab:code_classification} presents a comprehensive summary of the bespoke and non-bespoke techniques found in the generated code. The results reveal a rich repertoire of strategies spanning six categories: column encoding, query-support structures, scan and filter strategies, join operators, aggregation strategies, and low-level execution optimizations. At the storage level, the agent applies dictionary encoding to a quarter of all columns, physical sort orders to nearly 19\%, and compact representations such as scaled integers and 16-bit date offsets where applicable. On the query-execution side, bitmap semi-joins (73.7\% of queries) and inline fused aggregation (97.4\%) emerge as the most prevalent strategies, while hash indices for $O(1)$ join lookups and precomputed derived columns appear in 47.4\% and 36.8\% of queries, respectively. At the lowest level, pointer aliasing hints (\texttt{\_\_restrict}) and integer-scaled arithmetic are applied in roughly half of all queries, complemented by branchless range checks, branch hints, and software prefetching where beneficial. To illustrate how these bespoke and non-bespoke strategies manifest in concrete queries, we highlight two representative examples:

\paragraph{TPC-H Query~5.}
Q5 exploits the physical sort order of the \textit{orders} table by \textit{orderdate} to bound scans via \texttt{lower\_bound}, and uses a per-order range directory to touch only the relevant \textit{lineitem} slices, a strategy that benefits further from \textit{lineitems} being sorted by \textit{orderkey}. 
%The query reads only the few hot columns in a columnar layout, leveraging compact \texttt{int16} date offsets and integer-scaled prices for tight arithmetic. 
Denormalized attributes, i.e., \textit{lineitem.supp\_nationkey}, and \textit{discounted\_price}, are stored directly next to \textit{orders.cust\_nationkey}, rather than in separate tables, eliminates extra joins and redundant computations in the inner loops.
Region and nation lookups reduce to simple $O(1)$ array indexing via \textit{region.name\_to\_key} and dense key-to-nation arrays.

\paragraph{TPC-H Query~13.}
Q13 performs a focused two-column scan, touching only \textit{orders.custkey} and \textit{orders.comment} (plus \textit{customer.custkey} during aggregation).
Variable-length comments are accessed through a string arena (offsets + data) to enable fast offset-based substring checks. 
% Crucially, precomputed per-row masks (\textit{alpha\_mask}, \textit{bigram\_mask}) act as derived columns that quickly rule out non-matching rows, minimizing expensive substring searches.
Crucially, each order row carries two bitmasks: \textit{alpha\_mask}, recording which letter appear in comment, and \textit{bigram\_mask}, recording hashed character pairs, that cheaply rule out non-matching rows before any substring search is attempted.

\paragraph{Discussion of Results.}
While the Bespoke Agent draws from a large vocabulary of techniques, its selections are remarkably \emph{targeted}: for each query, it typically applies only a small, carefully chosen subset of strategies. This query-specific, and even column-specific, decision-making is precisely what distinguishes a bespoke engine from a general-purpose system, which must rely on broadly applicable but less optimal strategy choices. We provide the generated code alongside our Bespoke-OLAP framework.\footnote{Generated code: \url{https://github.com/DataManagementLab/BespokeOLAP_Artifacts}}

\vspace{-.5ex}
\subsection{The Synthesis Process}
\label{subsec:process}
We now shift our perspective from the \emph{output} of the Bespoke-OLAP agent to the \emph{process} by which the Bespoke Agent arrives at its implementations. Understanding the synthesis, including the development trajectory, tool usage patterns, and correctness behavior, is essential for assessing the practical viability of the approach.
\noindent\textbf{Development Trajectory and Speedups over Time.}
\begin{figure*}[!t]
\captionsetup{aboveskip=0.0ex,belowskip=0.0ex}
\captionsetup[subfigure]{aboveskip=0.0ex,belowskip=0.0ex}
    \centering
    \revisionimage{
    \includegraphics[width=\linewidth]{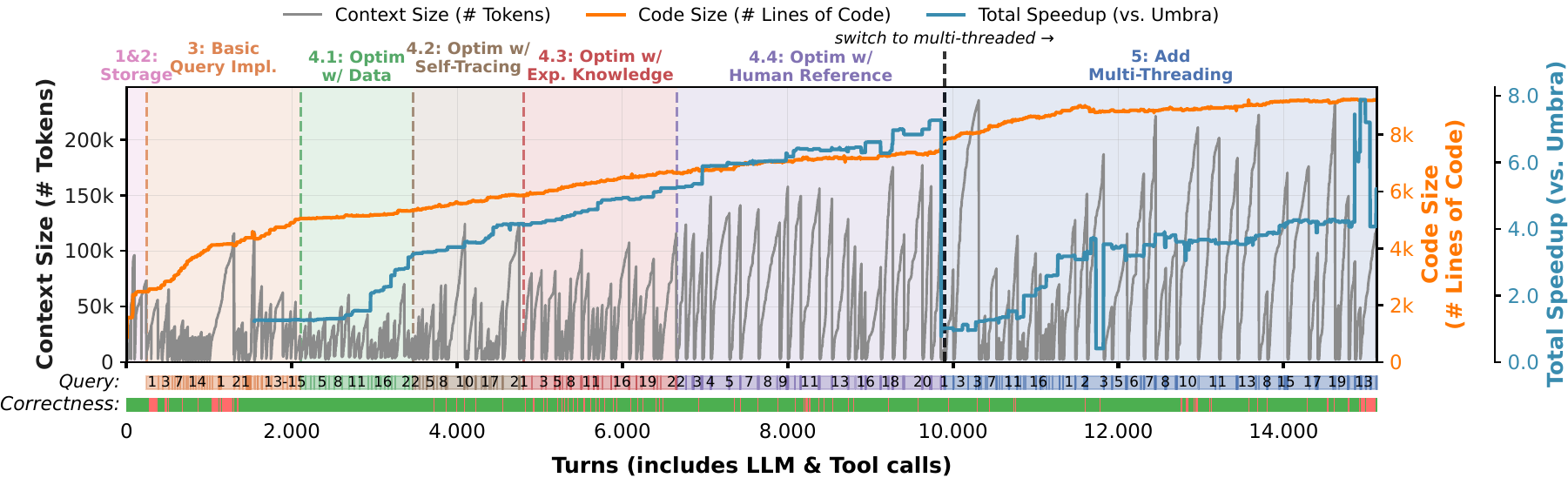}
    }
    \vspace{-1.5ex}
    \Description{Timeline of the Bespoke-TPCH synthesis process showing context size and lines of code over agent turns.}
    \caption{Development timeline of Bespoke-TPCH. The upper plot shows context size (left axis) and cumulative lines of code (right axis) over agent turns. Below the chart, the query currently being worked on is indicated, with green and red markers denoting correctness status. Compaction sessions periodically reduce the context size to prevent exceeding context limits.}
    \label{fig:timeline_turns}
    \vspace{-1.5ex}
\end{figure*}
\revision{
Figure~\ref{fig:timeline_turns} traces the synthesis process of Bespoke-TPCH from start to finish.
Over the course of approximately 15,000 agent turns, the agent progressed through all stages, starting from 5,700 lines of code and an initial speedup of 2.34$\times$ over DuckDB, and growing to a final engine of approximately 11,600 lines of C++, of which more than 3,000 lines are tracing infrastructure (hidden behind \texttt{\#ifdef} directives).
Occasional regressions (e.g. turn 11,600) during optimization were quickly detected and resolved via rollbacks. Context demands grew substantially in later stages, \change{peaking at over 100,000}{peaking at over 150,000} tokens, with regular compaction preventing context limit violations.}

\paragraph{Synthesis Cost.}
\revision{
The total cost of synthesizing Bespoke-TPCH amounted to \$276.44 in API costs (including \$251.02 for the optimization stages), while Bespoke-CEB required \$303.25 (\$259.26 for optimization).\footnote{These costs reflect the Claude Sonnet~4.7 pricing at the time of writing.}
The required wall-clock time ranges from 10-30 hours (including MT), depending on the LLM usage tier, rate limits, and queuing time.
}
Given that the resulting engines deliver order-of-magnitude speedups over DuckDB on their respective workloads, these one-time synthesis costs compare favorably to the cumulative savings in query execution time, particularly for workloads that are executed repeatedly over extended periods.

\vspace{-.5ex}
\section{Related Work}
\vspace{-0.5ex}
\label{sec:related_work}

\paragraph{Manual Specialization.}
A large body of research shows that specialization yields substantial performance gains. \change{Early analytical systems exposed the cost of generality, particularly the overhead of tuple-at-a-time execution and interpretation layers.  Vectorized and data-centric processing, as in MonetDB~/~X100~\cite{boncz2005monetdb}, mitigated these costs of generality by adapting query execution to data-intensive OLAP workloads. However, these are adaptations targeting only a workload class, not specific queries. Beyond query execution, the community has repeatedly demonstrated that system-level specialization can yield gains. Column stores and main-memory systems such as HyPer~\cite{kemper2011hyper} demonstrate the benefits of restricting generality improves performance. However, these systems are also specialized at the workload-class level rather than the individual-workload level and require substantial manual engineering effort. More extreme examples, such as TigerBeetle~\cite{tigerbeetle2024}, show that purpose-built systems for narrowly defined workloads can outperform general-purpose databases by orders of magnitude.  Yet, constructing a fully bespoke}{Vectorized and data-centric processing, as in MonetDB~/~X100~\cite{boncz2005monetdb}, mitigated the costs of generality by adapting query execution to data-intensive OLAP workloads, and column stores and main-memory systems such as HyPer~\cite{kemper2011hyper} further demonstrate that restricting generality improves performance. However, these systems specialize at the workload-class level rather than the individual-workload level; constructing a fully bespoke} engine for a specific deployment remained a manual engineering effort until this paper.

\paragraph{Query Compilation.}
Query compilation further advanced specialization by generating executable code tailored to individual queries, eliminating interpretation overhead and enabling aggressive inlining and loop fusion. 
HyPer~\cite{neumann2011efficiently,kemper2011hyper}, LegoBase~\cite{chafi2014building}, and DBToaster~\cite{ahmad2012dbtoaster} demonstrated the effectiveness of compiled query pipelines, and modern engines such as DuckDB~\cite{raasveldt2019duckdb} and Umbra~\cite{DBLP:conf/cidr/NeumannF20} adopt similar principles. \change{Despite these advances, query compilation remains restricted to the execution layer. Moreover, query compilation targets general SQL and does not optimize for a small set of queries.}{Despite these advances, query compilation remains restricted to the execution layer and targets general SQL rather than a small, fixed query set.}

\paragraph{DBMS Tuning.}
A dominant form of deployment-specific optimization is \emph{physical database design}, which selects workload-driven data structures such as indexes, clustering, partitioning, and materialized views. 
\change{These decisions are often made manually by DBAs using workload traces and execution plans, because the design space is large and alternatives interact (e.g., overlapping indexes and maintenance trade-offs).}{}
\change{Extensive work has automated these decisions within fixed DBMS architectures. Early AutoAdmin-style tools introduced cost-driven index selection and \emph{what-if} analysis for hypothetical structures~\cite{chaudhuri1997index,chaudhuri1998whatif}. Subsequent systems integrated indexes with materialized views and partitioning~\cite{agrawal2000views,agrawal2004partitioning}, and commercial tools such as SQL Server's Database Engine Tuning Advisor automate workload-driven physical design under resource constraints~\cite{agrawal2005dta}.}{Extensive work has automated these decisions, from early cost-driven index selection and \emph{what-if} analysis~\cite{chaudhuri1997index,chaudhuri1998whatif} to integrated views, partitioning, and commercial tuning advisors~\cite{agrawal2000views,agrawal2004partitioning,agrawal2005dta}.}
\change{A complementary line of work focuses on configuration tuning and self-driving databases~\cite{pavlo2017selfdriving,DBLP:conf/cidr/HilprechtBBEHKR20}, where systems such as OtterTune~\cite{vanaken2017automatic} apply machine learning to optimize system parameters. Other work incorporates learned models into core components, including learned indexes~\cite{kraska2018learned}, learned cardinality estimation~\cite{kipf2018learned}, and learned cost models~\cite{marcus2019neo}.}{A complementary line of work focuses on configuration tuning and self-driving databases~\cite{pavlo2017selfdriving,DBLP:conf/cidr/HilprechtBBEHKR20,vanaken2017automatic}, as well as learned models for indexes, cardinality estimation, and cost models~\cite{kraska2018learned,kipf2018learned,marcus2019neo}.}

\change{These methods can improve performance but are confined to a fixed architecture, optimizing only configurations and physical structures without altering the engine itself. In contrast, synthesizing a workload-specific DBMS requires exploring alternative system implementations, enabling deeper, architecture-level optimizations tailored to the workload.}{These methods can improve performance but are confined to a fixed architecture, whereas synthesizing a workload-specific DBMS enables deeper, architecture-level optimizations.}

\paragraph{LLM-Driven Synthesis.}
Large language models enable automated generation of nontrivial software artifacts~\cite{chen2021evaluating}, and tool-augmented agents can iteratively refine code using execution feedback~\cite{yao2022react}.
\change{These capabilities have begun to influence data management.
CodexDB~\cite{trummer2022codexdb}, for example, generates query-processing code from SQL queries and natural-language instructions by decomposing queries into plan steps and synthesizing executable implementations. 
Follow-up systems such as GPT-DB~\cite{trummer2023gptdb} generate customizable SQL processing code in general-purpose languages and support interactive control over execution behavior.}{CodexDB~\cite{trummer2022codexdb} generates query-processing code from SQL queries and natural-language instructions, and GPT-DB~\cite{trummer2023gptdb} generates customizable SQL processing code.}
However, neither aims for a bespoke design.
\change{More broadly, program synthesis and empirical autotuning systems such as FFTW~\cite{frigo1998fftw} and ATLAS~\cite{whaley2001automated} demonstrate that high-performance implementations can be generated automatically by exploring design alternatives.
However, existing LLM- and synthesis-based approaches target individual programs, kernels, or query-specific processing logic. 
They do not construct coherent systems with tightly coupled storage, execution, and architectural components. 
Database engines exhibit strong cross-layer dependencies, storage layouts constrain operators, execution depends on data organization, and optimizations introduce system-wide trade-offs that require structured construction and optimization across interacting subsystems.}{Program synthesis systems such as FFTW~\cite{frigo1998fftw} and ATLAS~\cite{whaley2001automated} similarly generate high-performance implementations automatically. Existing approaches target individual programs, kernels, or query-specific logic, however, and do not construct coherent systems with tightly coupled storage, execution, and architectural components, whereas database engines exhibit strong cross-layer dependencies requiring structured construction and optimization across interacting subsystems.}

\vspace{-0.5em}
\section{Conclusion and Future Work}
\label{sec:conclusion}
\revision{In this paper, we presented \emph{Bespoke OLAP}, a fully autonomous synthesis pipeline that leverages LLM-guided code generation to produce complete, workload-specific database engines from scratch. \change{On TPC-H (SF20), the synthesized engine achieves \emph{11.78$\times$ lower total runtime} and a \emph{16.40$\times$ median per-query speedup} over \emph{DuckDB v1.4.1}; on CEB (SF2), it achieves \emph{9.76$\times$} and \emph{4.66$\times$}, respectively, with improvements consistent across all query templates.}{On TPC-H (SF20), it achieves \SpeedupSTDuckTPCH{} over DuckDB v1.4.1 and \SpeedupSTUmbraTPCH{} over Umbra; on CEB (SF2), it achieves \SpeedupSTDuckCEB{} and \SpeedupSTUmbraCEB{}, respectively.} These results indicate that a substantial fraction of the overhead in modern analytical systems arises from their requirement to remain general-purpose, and that LLM-guided synthesis can systematically eliminate this cost on a per-workload basis. Because each engine is synthesized within hours at a cost of only a few dollars, per-workload specialization, which historically was reserved for exceptional, high-value deployments, can plausibly become routine for future DBMS-systems.}

For future directions, there are many interesting challenges and opportunities. \change{Our current pipeline targets in-memory OLAP to isolate the impact of workload-specific specialization from I/O scheduling. Extending the approach to disk-resident workloads will require synthesizing bespoke storage layouts and access paths. synchronization, and NUMA-aware data placement, and raises the open question of whether LLM agents can reliably generate correct concurrent code at this level of complexity.}{While our current submission already supports single-threaded disk-based execution, the achieved speedups do not yet reach the level of the in-memory setting. Closing this gap will require synthesizing bespoke buffer pool management, eviction policies, and block-aligned page layouts tailored to the known query access pattern. We consider this an important direction, though one that introduces a substantially larger synthesis problem which we will tackle in the future.} \change{Incorporating}{Beyond disk-resident workloads, incorporating} lightweight cost models tailored to LLM-generated code could further reduce synthesis time by pruning unpromising optimization alternatives early, particularly for join ordering. Finally, extending bespoke synthesis beyond pure OLAP to other workloads such as OLTP introduces consistency and isolation guarantees as a fundamentally new dimension of the synthesis problem.

\vspace{-1ex}
\begin{acks}
  \vspace{-.5ex}
\small
This research was partially funded by the state of Hesse as part of the NHR program (NHR4CES), the LOEWE Spitzenprofessur of the state of Hesse (III 5-519/05.00.003-(0005)), the project "hessian.AI AISC" (Grant No. 01IS22091) by the Federal Ministry of Education and Research (BMBF), as well as by the Deutsche Forschungsgemeinschaft (DFG, German Research Foundation) under Germany's Excellence Strategy (EXC3057/1 “Reasonable Artificial Intelligence”, Project No. 533677015). 
We also thank DFKI Darmstadt and hessian.AI for their support.
\par
\end{acks}

\bibliographystyle{ACM-Reference-Format}
\bibliography{bib}

\end{document}